\documentstyle[11pt]{article}
\newtheorem{theorem}{Theorem}
\newtheorem{lemma}{Lemma}

\newcommand {\bg} {\mbox{\boldmath $g$}}

\newcommand {\bv} {\mbox{\boldmath $v$}}

\newcommand {\bx} {\mbox{\boldmath $x$}}
\newcommand {\by} {\mbox{\boldmath $y$}}
\newcommand {\bz} {\mbox{\boldmath $z$}}

\newcommand {\bG} {\mbox{\boldmath $G$}}

\newcommand {\bX} {\mbox{\boldmath $X$}}

\newcommand {\bvm} {\mbox{\boldmath $v_m$}}
\newcommand {\bvmtag} {\mbox{\boldmath $v_{m'}$}}
\newcommand {\bvtag} {\mbox{\boldmath $v'$}}
\newcommand {\bvj} {\mbox{\boldmath $v_j$}}
\newcommand {\bVj} {\mbox{\boldmath $V_j$}}

\newcommand {\bVjtag} {\mbox{\boldmath $V_j'$}}

\newcommand {\bVjtagaim} {\mbox{\boldmath $V_j''$}}

\newcommand {\bvt} {\mbox{\boldmath $v_t$}}
\newcommand {\bvzero} {\mbox{\boldmath $v_0$}}
\newcommand {\bvzerot} {\mbox{\boldmath $v_0^{t}$}}
\newcommand {\bvjplusi} {\mbox{\boldmath $v_{j+i}$}}
\newcommand {\bvtagjplusi} {\mbox{\boldmath $v'_{j+i}$}}

\newcommand {\bvjplusone} {\mbox{\boldmath $v_{j+1}$}}

\newcommand {\bvzerojplusi} {\mbox{\boldmath $v_{0}^{j+i}$}}
\newcommand {\bvjplusKpluslminusone} {\mbox{\boldmath $v_{j+K+l-1}$}}

\newcommand {\bVjupi} {\mbox{\boldmath $V_j^{(i)}$}}
\newcommand {\bVjupiminusone} {\mbox{\boldmath $V_j^{(i-1)}$}}
\newcommand {\bVjupiplusone} {\mbox{\boldmath $V_j^{(i+1)}$}}
\newcommand {\bVjupone} {\mbox{\boldmath $V_j^{(1)}$}}
\newcommand {\bVjupM} {\mbox{\boldmath $V_j^{(M)}$}}

\newcommand {\bolVj} {\mbox{\boldmath $\overline{V}_j$}}
\newcommand {\bolVjtag} {\mbox{\boldmath $\overline{V}_j'$}}
\newcommand {\bolVji} {\mbox{\boldmath $\overline{V}_j^{(i)}$}}

\newcommand {\bum} {\mbox{\boldmath $u_m$}}
\newcommand {\bumtag} {\mbox{\boldmath $u_{m'}$}}
\newcommand {\buj} {\mbox{\boldmath $u_j$}}
\newcommand {\buzero} {\mbox{\boldmath $u_0$}}
\newcommand {\butminusj} {\mbox{\boldmath $u_{t-j}$}}
\newcommand {\bujplusone} {\mbox{\boldmath $u_{j+1}$}}
\newcommand {\bujpluslminusone} {\mbox{\boldmath $u_{j+l-1}$}}
\newcommand {\bujplusl} {\mbox{\boldmath $u_{j+l}$}}

\newcommand {\bgi} {\mbox{\boldmath $g_i$}}
\newcommand {\bgj} {\mbox{\boldmath $g_j$}}
\newcommand {\bgitilde} {\mbox{\boldmath $\tilde{g_i}$}}
\newcommand {\bgjtilde} {\mbox{\boldmath $\tilde{g_j}$}}

\newcommand {\bxm} {\mbox{\boldmath $x_m$}}
\newcommand {\bxtag} {\mbox{\boldmath $x'$}}
\newcommand {\bxone} {\mbox{\boldmath $x_1$}}
\newcommand {\bxtwo} {\mbox{\boldmath $x_2$}}
\newcommand {\bxmtag} {\mbox{\boldmath $x_{m'}$}}

\newcommand {\bolx} {\mbox{\boldmath $\overline{x}$}}

\newcommand {\bytag} {\mbox{\boldmath $y'$}}
\newcommand {\bYj} {\mbox{\boldmath $Y_j$}}

\newcommand {\bzero} {\mbox{\boldmath $0$}}

\newcommand {\bGj} {\mbox{\boldmath $G_j$}}
\newcommand {\bGjt} {\mbox{\boldmath $G_j^t$}}
\newcommand {\bGzerojplusi} {\mbox{\boldmath $G_0^{j+i}$}}
\newcommand {\bGonejplusi} {\mbox{\boldmath $G_1^{j+i}$}}
\newcommand {\bGKminusonejplusi} {\mbox{\boldmath $G_{K-1}^{j+i}$}}

\newcommand {\bXtag} {\mbox{\boldmath $X'$}}

\newcommand{\calC}{{\cal C}}

\newcommand{\calQ}{{\cal Q}}

\newcommand{\calV}{{\cal V}}

\newcommand{\calX}{{\cal X}}
\newcommand{\calY}{{\cal Y}}

\begin{document}

\setlength{\baselineskip}{1.5\baselineskip}

\title{Competitive Minimax Universal Decoding for Several Ensembles of Random Codes
\thanks{This research was supported by the Israel Science Foundation (ISF), grant no.\ 223/05.} }


\author{Yaniv Akirav and Neri Merhav \\
Department of Electrical Engineering \\
Technion - Israeli Institute of Technology \\
Technion City, Haifa 32000, Israel\\
Emails:[yaniva@tx,merhav@ee].technion.ac.il
}


%


\maketitle

\begin{abstract}
Universally achievable error exponents pertaining to certain
families of channels (most notably, discrete memoryless channels (DMC's)), and various
ensembles of random codes, are studied by combining the competitive minimax approach, proposed by Feder and Merhav, with Chernoff bound and Gallager's techniques for the analysis of error exponents. In particular, we derive a single--letter expression for the largest, universally achievable fraction $\xi$ of the optimum error exponent pertaining to the optimum ML decoding.
Moreover, a simpler single--letter expression for a lower bound to $\xi$ is presented. To demonstrate the tightness of this lower bound, we use it to show that $\xi=1$, for the 
binary symmetric channel (BSC), when the random coding distribution is
uniform over:
(i) all codes (of a given rate), and (ii) all linear codes, in agreement with well--known results. 
We also show that $\xi=1$ for the uniform ensemble of systematic 
linear codes, and for that of time--varying convolutional 
codes in the bit-error--rate sense.
For the latter case, we also show how the corresponding universal decoder can be efficiently
implemented using a slightly modified version of the Viterbi algorithm which employs two trellises.\\

\noindent
{\bf Index Terms:} error exponent, universal decoding, generalized likelihood ratio test, channel uncertainty, competitive minimax, Viterbi algorithm, maximum mutual information decoding.

\end{abstract}


%


\clearpage

\section{Introduction}

In many real--life situations, encountered in digital coded communication systems, channel variability and uncertainty prohibit the use of the optimum maximum likelihood (ML) decoder, and so, universal decoders, independent of the unknown channel parameters, are sought.

The topic of universal coding and decoding for unknown channels has received considerable attention in the last three decades.
In \cite{G75}, Goppa offered the \textit{maximum mutual information} (MMI) decoder, which decides in favor of the code vector with maximum empirical mutual information with the channel output. Goppa showed that for DMC's, MMI decoding achieves capacity.
Csisz\'ar and K\"orner \cite{CK81} also explored the universal decoding problem for DMC's with finite input and output alphabet. They showed that the random coding error exponent associated with a uniform random coding distribution over a type class achieves the optimum error exponent.
Csisz\'ar \cite{C82} proved that for any channel within the class of DMC's with additive noise, and the uniform random coding distribution over linear codes, the optimum error exponent is achievable by a decoder minimizing the noise empirical entropy, universally for all the channels in the class.
Ziv \cite{Ziv85} explored the universal decoding problem for finite state channels with finite input and output alphabets, for which the next channel state is a deterministic (but unknown)  function of the channel current state and current inputs and outputs. For codes governed by a uniform random coding over a given set, he proved that a decoder based on the Lempel-Ziv algorithm asymptotically achieves the error exponent associated with ML decoding.
In \cite{LZ98}, Ziv and Lapidoth proved that the latter decoder is universal for a wider class of finite--state channels.
In \cite{FL98}, Feder and Lapidoth found sufficient conditions for families of channels, to have universal decoders that asymptotically achieve the random coding error exponent associated with ML decoding.

Universal coding and decoding were explored also with regard to the generalized likelihood ratio test (GLRT). In this approach, each message is scored according to the maximum likelihood (over the parameter space) of the channel output vector given the message, and a decision is made in favor
of the message that attains the highest maximum likelihood. Although provably optimum in certain asymptotic situations \cite{ZZM92}, \cite[p.\ 165, Theorem 5.2]{CK81}, there are cases where the GLRT is strictly suboptimum \cite[Sect.\ III, pp.\ 1754--1755]{LZ98}, \cite[Appendix]{FM02}.

The competitive minimax criterion, first presented in \cite{FM02}, is an attempt for a general methodological approach to the problem of universal decoding. According to this approach, the criterion is the minimum (over all decision rules) of the maximum (over all channels in the family) of the ratio between the error probability associated with a given channel and given decision rule, and the error probability of the ML decoder for that channel, raised to some power $\xi\in[0,1]$ (cf.\ eq.\ (\ref{decision_criterion}) below).
The largest power $\xi=\xi^*$ such that the value of this minimax ratio does not grow exponentially with the block length, is the maximum universally achievable fraction of the ML error exponent.

The main contribution of this paper is in deriving a single--letter expression to $\xi^*$, in terms of the rate $R$ and a general random coding distribution, for fairly general families of channels and ensembles of random codes. While in previous works the universality was proved for certain channel models (e.g. finite--state channels, etc.) and random coding distributions (e.g. uniform distribution over a given type class, etc.), this work deals with general families of DMC's (cf. Sect. II) and general random coding distributions (cf.\ eq.\ (\ref{TBD definition})).
We should note that a similar technique can be used to broaden the result for $\xi^*$ to other channel families, e.g. Markov channels, finite state channels, etc.\\
In addition, a single--letter expression for a lower bound to $\xi^*$ is presented, which is simpler to work with, and is believed to be tight. This lower bound is true also for random coding distribution over ensembles of linear code and systematic linear codes. The tightness of this lower bound is demonstrated for the case of the BSC. For this model, we show that $\xi^*=1$, when the random coding distribution is uniform over all codes and over all linear codes, in agreement with well--known results.
We also show that $\xi^*=1$ for the ensemble of systematic linear codes, and for that of time--varying convolutional codes in the bit-error--rate sense.
Using the fact that in the case of the BSC, the minimax decoding metric degenerates to a simpler metric, we propose an efficient implementation based on a slightly modified version of the Viterbi algorithm.

The outline of the paper is as follows. In Section II, we establish the notation that will be used throughout the paper and provide a formal definition of the universal decoding problem. In Section III, the main results are stated and discussed. Section IV contains a detailed proof of the single--letter expression for $\xi^{*}$ will be provided.
In Section V, the tightness of the lower bound to $\xi^{*}$ is demonstrated for the case of the BSC with an unknown crossover probability. In Section VI, we prove that for the ensemble of time-varying convolutional codes and the BSC with an unknown crossover probability, the minimax decoder achieves the same bit error exponent as the ML decoder, which is used when the parameter is known.

\section{Notation and Problem Definition}
Throughout this paper, scalar random variables (RV's) will be denoted by capital letters, their sample values will be denoted by the respective lower case letters, and their alphabets will be denoted by the respective calligraphic letters.
A similar convention will apply to random vectors of dimension $N$ and their sample values,
which will be denoted with same symbols in the bold face font.
The set of all $N$--vectors with components taking values in a certain alphabet,
will be denoted as the same alphabet superscripted by $N$.\\
Information theoretic quantities like entropies, conditional entropies, and mutual informations,
will be denoted following the usual conventions of the information theory literature, e.g., $H(X)$, $H(X|Y)$, $I(X;Y)$, and so on. With a slight abuse of notation, when we wish to emphasize the dependence of the entropy on the underlying probability distribution $P$, we denote it by $H(P)$.\\
The mutual information between the input and the output of the channel\\ $\left\{P_\theta \left(y|x \right),x\in {\cal X},y\in {\cal Y}\right\}$, when the input is governed by $Q$, will be denoted by
\begin{equation}
I_{\theta}\left(Q\right)=
\sum_{x\in {\cal X}}\sum_{y\in {\cal Y}}
Q\left(x\right)P_\theta \left(y|x \right)
\ln
\frac
{P_\theta \left(y|x \right)}
{
\sum_{x\in {\cal X}}
Q\left(x\right)
P_{\theta}\left(y|x\right)
},
\end{equation}
and the capacity of the channel will be denoted by
$
C_{\theta}=\max_{Q}I_{\theta}\left(Q\right)
$.\\
The number of occurrences of a letter $a\in{\cal X}$ in a vector $\bx\in{\cal X}^N$ will be denoted by $N_{\bx}(a)$. The empirical distribution of $\bx$ will be denoted by $P_{\bx}=\left\{P_{\bx}(a)=N_{\bx}(a)/{N},\ a\in{\cal X}\right\}$.
The type class of $\bx$ is defined as $T_{\bx}=\left\{\bxtag:P_{\bxtag}=P_{\bx}\right\}$ and $H_{\bx}(X)=-\sum_{a\in{\cal X}}
	P_{\bx}\left(a\right)\ln{P_{\bx}\left(a\right)}$ will denote the entropy of a random variable (RV) $X$, with distribution $P_{\bx}$.
Similarly, the number of occurrences of a letter pair $\left(a,b\right)\in{\cal X}\times{\cal Y}$ in the vector pair $(\bx,\by)$ will be denoted by $N_{\bx\by}(a,b)$,
$
P_{\bx\by}=\left\{P_{\bx\by}\left(a,b\right)=
	N_{\bx\by}(a,b)/{N},\ \left(a,b\right)\in{\cal X}\times{\cal Y}\right\}
$
will denote the joint empirical distribution of $(\bx,\by)$,
$T_{\bx\by}=\left\{\bxtag,\bytag:P_{\bxtag,\bytag}=P_{\bx\by}\right\}$ 
will stand for the joint type class of $(\bx,\by)$, and
$H_{\bx\by}(X,Y)=-\sum_{a,b\in{\cal X}\times{\cal Y}}
P_{\bx\by}\left(a,b\right)
\ln{P_{\bx\by}\left(a,b\right)}$
will denote the joint entropy of RV's $(X,Y)$ with joint distribution $P_{\bx\by}$.
We will use $T_{\bx|\by}=\left\{\bxtag:P_{\bxtag\by}=P_{\bx\by}\right\}$ to denote the conditional type class of $\bx$ given $\by$, 
$
P_{\bx|\by}\left(a|b\right)=
	N_{\bx\by}(a,b)/N_{\by}(b),\ \left(a,b\right)\in{\cal X}\times{\cal Y},
$
to denote the conditional empirical distribution related to $\left(a,b\right)\in{\cal X}\times{\cal Y}$,
and $H_{\bx\by}(X|Y)=-\sum_{a,b\in{\cal X}\times{\cal Y}}
P_{\bx\by}\left(a,b\right)
\ln P_{\bx|\by}\left(a|b\right)$ 
to denote the conditional entropy of $X$ given $Y$, induced by the joint distribution $P_{\bx\by}$.
The empirical mutual information between RV's $X$ and $Y$ with joint distribution $P_{\bx\by}$ will be denoted by $I_{\bx\by}(X;Y)=H_{\bx}(X)-H_{\bx\by}(X|Y)$.\\
The expectation of a function $F(X,Y)$, where $X$ and $Y$ are RV's distributed according to the empirical distribution of $\bx$ and $\by$, will be denoted by $$\hat{E}_{\bx\by}\left\{F(X,Y)\right\}=\sum_{a\in{\cal X}}\sum_{b\in{\cal Y}}P_{\bx\by}(a,b)F(a,b).$$
The notation 
$
E_{Q}
	\left\{
		F(\bX)
	\right\}
$ will be used for the expectation of a function $F(\bX)$, where the random vector $\bX$ is governed by $Q$.\\
The Hamming distance between two vectors $\bx$ and $\by$ will be denoted by 
$d{(\bx,\by)}$, and its normalization by $N$ will be denoted by
$\delta{(\bx,\by)}$.
 For a finite set ${\cal A}$, $|{\cal A}|$ will stand for its cardinality.  
The divergence between two probability measures $P$ and $Q$ over an alphabet ${\cal U}$ will be denoted by 
$
D\left(P||Q\right)=\sum_{u\in {\cal U}}
P\left(u\right)\ln\frac{P\left(u\right)}{Q\left(u\right)}
$, 
where $0\ln 0$ and $0\ln \frac{0}{0}$ are defined as $0$, and
$P\ln\frac{P}{0}$ for $P > 0$ is defined as $\infty$.
For two positive sequences $\{A_N\}_{N\ge 1}$ and $\{B_N\}_{N\ge 1}$, the notation $A_N\stackrel{\cdot}{=} B_N$ will express the fact that $\{A_N\}_{N\ge 1}$ and $\{B_N\}_{N\ge 1}$ are of the same exponential order, i.e., $$\lim_{N\to\infty}\frac{1}{N}\ln \left(A_N/B_N\right)=0.$$
 
Consider a DMC with a finite input alphabet ${\cal X}$, a finite output alphabet ${\cal Y}$, and single letter transition probabilities $\left\{P_\theta \left(y|x \right),x\in {\cal X},y\in {\cal Y}\right\}$, where $\theta$ is an unknown parameter vector, taking values in some set $\Theta$. The channel is fed by an input vector of length $N$, $\bx\in {\cal X}^{N}$, and generates an output vector $\by\in {\cal Y}^{N}$ according to $P_\theta(\by|\bx)=\prod_{i=1}^N
P_\theta(y_i|x_i)$.
A rate-$R$ block code of length $N$ consists of $M=e^{NR}$ $N$--vectors $\bxm\in {\cal X}^{N}$, $0 \leq m \leq M-1$, representing $M$ different messages.
A decoder $\Omega$ is a partition of ${\cal Y}^N$ into $M$ regions,
$\Omega_0,\Omega_1,\ldots,\Omega_{M-1}$, such that if $\by$ falls into $\Omega_m$, a decision is made in favor of message $m$.

Given a code ${\cal C}$, the competitive minimax criterion \cite{FM02} is defined as
\begin{equation}
\label{decision_criterion}
S_N \stackrel{\Delta}{=}
\min_\Omega\max_{\theta \in \Theta}
		\left\{
				\frac
					{P_{E}\left(\Omega|\theta\right)}
					{[{P_{E}}^{*}\left(\theta\right)]^\xi}
		\right\},\ \ \ 0\leq\xi\leq1,
\end{equation}
where $P_{E}\left(\Omega|\theta\right)=\frac{1}{M}\sum_{m=0}^{M-1}\sum_{\by \in \Omega_m^c}
P_{\theta}(\by|\bxm)$
is the error probability related to a decoder $\Omega$ for a given value of $\theta$, and $P_{E}^{*}\left(\theta\right)=\min_\Omega P_{E}(\Omega|\theta)$ is the ML decoding error probability when $\theta$ is known. 

The ratio 
$
					{P_{E}\left(\Omega|\theta\right)}
					/
					{[{P_{E}}^{*}\left(\theta\right)]^\xi}
$
designates the loss in error probability, caused by using a universal decoder which is ignorant of $\theta$, relative to the optimal ML decoding for that $\theta$. The parameter $\xi$ can be interpreted as the fraction of the optimal error exponent to which the universal decoder error exponent is compared. In order to minimize this loss uniformly over all $\Theta$, a decoder $\Omega$ which minimizes the worst case of that ratio (i.e., its maximum), is sought.

As $S_N$ addresses the ratio between the error probabilities, it corresponds to the difference between the error exponents related to these errors. It is well known that for most channels, the decoding error decays exponentially with the block length $N$. Therefore, if the value of $S_N$, for a decision rule $\Omega$ achieved by (\ref{decision_criterion}), grows sub--exponentially with $N$, i.e., $\lim_{N\to\infty}\frac{1}{N}\ln S_N=0$, it means that, uniformly over $\Theta$, the error probability associated with $\Omega$ decays with an exponential rate which is at least a fraction $\xi$ of the error exponent rate of ${P_{E}}^{*}\left(\theta\right)$.

In \cite{FM02}, the following decision rule has been shown to be asymptotically optimal in the minimax sense for a given $\xi$: 
\begin{equation}
\label{explicit_decision_rule_2}
\Omega_{m} =
	\left\{
	\by |
			f(\bxm,\by) 
			\geq
			f(\bxmtag,\by),\ \forall{m'}\neq m
	\right\}
\end{equation}
with ties broken arbitrarily, where
\begin{eqnarray}
f(\bx,\by) 
& \stackrel{\Delta}{=} &
\max_{\theta \in \Theta}	f_{\theta}(\bx,\by),
\\
f_{\theta}(\bx,\by)
& \stackrel{\Delta}{=} &
		\frac{1}{N}
					\ln\:{P_\theta\left(\by|\bx\right)}+
					\xi E^{*}(\theta),
\end{eqnarray}
and $E^{*}(\theta)$ stands for the asymptotic exponent associated with
$P_{E}^{*}\left(\theta\right)$.
A decoder $\Omega$, defined by ($\ref{explicit_decision_rule_2}$), will be called the \textit{minimax decoder} hereafter.

A natural question that may arise, at this point, is with regard to the choice of the free parameter $\xi$. As mentioned above, the main guideline proposed in \cite{FM02} is to seek the maximum value $\xi^*$ of $\xi$ such that $S_N$ would still grow sub--exponentially with $N$.

In the random coding regime, the error probabilities at the numerator and the denominator of (\ref{decision_criterion}) are replaced by the corresponding average error probabilities, i.e.,
\begin{equation}
\label{average_decision_criterion}
\overline{S}_N
\stackrel{\Delta}{=}
\min_\Omega\max_{\theta \in \Theta}\left\{\frac
			{\overline{P}_{E}\left(\Omega|\theta\right)}
			{[{\overline{P}_{E}}^{*}\left(\theta\right)]^\xi}\right\}
\end{equation}
and the decoder (\ref{explicit_decision_rule_2}) is used, with $E^{*}(\theta)$ being replaced by $E_{r}^{*}(\theta)$, the random coding error exponent associated with
$\overline{P}_{E}^{*}\left(\theta\right)$.

The main purpose of this paper is to translate the above--mentioned guideline for the choice of $\xi$ into a concrete single--letter formula for the random coding regime.

\section{Statement of Results}

In this section, by evaluating the exponential order of $\overline{S}_N$, we derive a formula for $\xi^*$, the largest value of $\xi$ for which $\overline{S}_N$ is sub--exponential in $N$. 
Moreover, an expression for the lower bound to $\xi^*$ is also derived, and its tightness is demonstrated for the BSC model and for several ensembles of random codes.

\subsection{General codes}
We begin with a few definitions. For every positive integer $N$,
let $Q_N$ be a random coding distribution for $N$--vectors, of the following form:
\begin{equation}
\label{TBD definition}
Q_N(\bx)=\frac{Q_N(T_{\bx})}{|T_{\bx}|},
\end{equation}
i.e., uniform distribution for all the vectors within the same type class.
Of course, $$\sum_{T_{\bx}}Q_N(T_{\bx})=1.$$ Now, let
$$\Delta_N(P_{\bx})=-\frac{1}{N}\ln Q_N(T_{\bx}),$$
and let $\Delta_N^*(P)$ be an extension of the function $\Delta_N(P_{\bx})$ 
that is defined over the continuum of probability distributions over $\calX$
(rather than just the set of rational probability distributions with denominator $N$). 
We next define the class $\calQ$ of sequences of random coding distributions $\{Q_N\}$ as follows: A sequence of random coding distributions $\{Q_N\}_{N\ge 1}$ is said to belong to the
class $\calQ$ if there exists such an extension $\Delta_N^*(P)$ that converges, as $N\to\infty$, to a certain non--negative functional $\Delta^*(P)$, uniformly over all probability distributions $\{P\}$ over $\calX$.\\
It is easy to see that the class $\calQ$ essentially covers all random coding distributions that
are customarily used (and much more). 
In particular, to approximate a random coding distribution which is uniform within 
a small neighborhood of one type class -- corresponding to a probability distribution $P_0$, and which vanishes elsewhere, we set $\Delta^*(P)=0$ for every $P$ in that neighborhood of $P_0$, and $\Delta^*(P)=\infty$ elsewhere. For the case where $Q$ is i.i.d., $\Delta^*(P)=D(P\|Q)$. In particular, if $Q(\bx)=1/|\calX|^N$ for all $\bx\in\calX^N$, then $\Delta^*(P)=\ln|\calX|-H(P)$.

Given a joint distribution $P_{XY}$, a real $\alpha$, and a value of $\theta \in \Theta$, let
\begin{eqnarray}
\label{A_theta}
A(\theta,\alpha,P_{XY})
& \stackrel{\Delta}{=} &
						 		I(X;Y)
								+\Delta^{*}\left(\sum_{b\in\calY}P_{Y}(b)P_{X|Y}(\cdot|b)\right)
						 		-{\alpha}E
												\ln{
													P_\theta(Y|X)
													},
\end{eqnarray}
where $E\{\cdot\}$ is the expectation and $I(X;Y)$ is the mutual information w.r.t.\ a generic
joint distribution $P_{XY}(a,b)=P_Y(b)P_{X|Y}(a|b)$ of the RV's $(X,Y)$.

Next, for distributions $P_{Y}$,$P_{X|Y}$ and $P_{X'|Y}$, two parameters $\theta, \theta' \in \Theta$, and reals $0\leq \rho\leq 1$ and $s \geq 0$, define:
\begin{equation}
\label{B_definition}
	B(\theta,\theta',P_{Y},P_{X|Y},P_{X'|Y},s,\rho)
	\stackrel{\Delta}{=}
						A(\theta,1-s \rho,P_{XY})
						+\rho \cdot 
 						A(\theta',s,P_{X' Y})
						-H(Y),
\end{equation}
where $H(Y)$ is the entropy of $Y$ induced by $P_Y$.
Finally, let
\begin{eqnarray}
\label{xi_star_definition - 1}
\xi^{*}(R)
& = &
\min_{P_{X Y}}
\min_{\theta' \in \Theta}
\max
\Biggl\{
\min_{\theta \in \Theta}			
\max_{\stackrel{0 \leq \rho \leq 1}{0 \leq s \leq 1/\rho}}
\min_{P_{X' | Y}}
\frac{B(\theta,\theta',P_{Y},P_{X| Y},P_{X' | Y},s,\rho)-\rho R}
			{(1-\rho s) E_r^{*}(\theta)+\rho s E_r^{*}(\theta')}
,
\nonumber\\
\label{xi_star_definition - 2}
& &
\ \ \ \ \ \ \ \ \ \ \ \ \ \ \ \ \ \ \ \ \ \ \ 
\max_{\theta \in \Theta}
\max_{\stackrel{0 \leq \rho \leq 1}{s  \geq 1/\rho}}
\min_{P_{X' | Y}}
\frac{B(\theta,\theta',P_{Y},P_{X| Y},P_{X' | Y},s,\rho)-\rho R}
			{(1-\rho s) E_r^{*}(\theta)+\rho s E_r^{*}(\theta')}
\Biggr\}
\nonumber\\
\end{eqnarray}

Our main result, in this section, is the following:
\begin{theorem}
Consider a sequence of ensembles of codes, where each codeword is drawn independently,
under a distribution $Q_N$, and the sequence $\{Q_N\}_{N\ge 1}$ is a member of the class $\calQ$.  Then,
\begin{enumerate}
\item For every $\xi \leq \xi^{*}\left(R\right)$, 
$\lim_{N\to\infty} \frac{1}{N}\ln\overline{S}_N \leq 0.$
\item There exists a sequence of encoders
$\left\{\calC_N\right\}_{N\ge 1}$
and minimax decoders
$\left\{\Omega_N\right\}_{N\ge 1}$ with $\xi = \xi^{*}\left(R\right)$,
for which:
			 $$\liminf_{N\to\infty}\left[
			 													-\frac{1}{N}\ln P_{E}\left(\Omega_N|\theta\right)
			 												\right]
			 												\ge
					\xi \cdot E^{*}(\theta)$$
uniformly over $\theta \in \Theta$.
\item For every $\xi > \xi^{*}\left(R\right)$, 
$\lim_{N\to\infty} \frac{1}{N}\ln\overline{S}_N > 0.$
\end{enumerate}
\end{theorem}

The proof of Theorem 1 appears in Section IV.

We now pause to discuss Theorem 1 and some of its aspects.\\	
The theorem suggests a conceptually simple strategy for universal decoding: Given $R$ and the sequence $\{Q_N\}_{N\ge 1}$, first, compute
$\xi^{*}\left(R\right)$ using eq.\ (\ref{xi_star_definition - 1}). This may require some non-trivial optimization procedures, but it has to be done only once. It should be mentioned that if closed--form analytic expression does not seem available, the computation can be carried out at least numerically, since this is a single--letter expression.
Once $\xi^{*}\left(R\right)$ has been computed, apply the minimax decoding rule with $\xi=\xi^{*}\left(R\right)$ and the theorem guarantees that the resulting random coding error exponent associated with the decoder is as specified in the second item of that theorem. Moreover, the third item of the theorem implies that in the random coding regime, $\xi^{*}\left(R\right)$ is the largest fraction of $E^{*}(\theta)$ that is uniformly achievable by a universal decoder.

As mentioned earlier, when $Q$ is uniform i.i.d.,
$\Delta^*(P)=\ln|\calX|-H(X)$ (where $X$ is governed by $P$),  and therefore
\begin{eqnarray}
\label{A_theta_for i.i.d.}
A(\theta,\alpha,P_{XY})
& = &
\ln|\calX|
-H(X|Y)
						 		-{\alpha}E
												\ln{
													P_\theta(Y|X)
													}
							.
\end{eqnarray}
This observation will be used in Section V which deals with the BSC model, as well as in Section A.1 of the Appendix (ensembles of linear and systematic linear codes), as they both assume a binary i.i.d. random coding distribution.
%

The theorem is interesting, of course, only when $\xi^{*}\left(R\right) >0$, which is the case in many situations, at least as long as $R$ is not too large. It should be pointed
out that the exponential rate $\xi^{*}\left(R\right)\cdot E^{*}(\theta)$,
guaranteed by Theorem 1, is only a lower bound to the real exponential rate (as the minimax criterion is aimed to consider all $\theta \in \Theta$), and that true
exponential rate, at some points in $\Theta$, might be larger.

As mentioned above, the exact formula for $\xi^{*}$, given in eq. (\ref{xi_star_definition - 1}), includes many optimizations and hence might be complicated for calculation. Therefore, we next present a simpler expression for a lower bound to $\xi^{*}$, denoted by $\xi_{LB}^{*}\left(R\right)$, which we believe is tight at least for several families of channels. Another motivation for presenting $\xi_{LB}^{*}\left(R\right)$ is that it holds also for ensembles of linear and systematic linear codes, as we will shall in the next subsection.
The expression for $\xi_{LB}^{*}\left(R\right)$ will be derived from $\xi^{*}\left(R\right)$ by: (i) avoiding the inner maximization between two terms in (\ref{xi_star_definition - 1}) by choosing the left term, and (ii) interchanging between the minimization over $P_{X| Y}$ and the maximization over $\lambda$ and $\rho$, i.e:
\begin{equation}
\label{xi star - lower bound}
\xi_{LB}^{*}\left(R\right)
\stackrel{\Delta}{=}
\min_{P_{Y}}
\min_{\theta \in \Theta}
\min_{\theta' \in \Theta}
\max_{\stackrel{0\leq\rho\leq1}{0\leq\lambda\leq 1/\rho}}
\min_{P_{X| Y}}
\min_{P_{X'|Y}}
		\frac{
				B(\theta,\theta',P_{Y},P_{X|Y},P_{X'|Y},\lambda,\rho)
				-\rho R
		}
		{
			(1- \lambda \rho) \cdot E_{r}^{*}(\theta)
			+\lambda \rho \cdot E_{r}^{*}(\theta')
		}.
\end{equation}
As $\xi_{LB}^{*}\left(R\right)$ is a lower bound to $\xi^{*}$, it is obvious to see that parts 1 and 2 of Theorem 1 hold for it as well.
%
\subsection{Linear codes}

We next provide a variation of $\xi_{LB}^{*}\left(R\right)$ for ensembles of linear codes and systematic linear codes. Prior to that, we first define these ensembles.
A linear code is defined by mapping each of the $M=2^K$ binary information (row) vectors $\bum,\ 0\leq m \leq M-1$, of length $K$, into its corresponding code (row) vector $\bvm$, of length $N$, in the following way:
$$\bvm=\bum \bG\oplus\bvzero,\ \ \ m=0, 1,\ldots,M-1,$$
where $\bG$ is a binary generator matrix of dimension $K\times N$ and $\bvzero$ is an additive vector of length $N$. The $\oplus$ operation denotes a summation modulo 2 and the multiplication between $\bum$ and $\bG$ is conducted over the field $GF(2)$.
A systematic linear code is defined in the same manner, with the restriction that the left $K\times K$ block of $\bG$ (the systematic part of $\bG$) forms the identity matrix (thus, the first $K$ bits of each code vector, $\bvm$, form the corresponding information vector, $\bum$).

We now consider a random coding distribution, which is i.i.d. over the ensemble of linear codes (or systematic linear codes), for which the elements of $\bG$ (or $\tilde{\bG}$, the non-systematic part of $\bG$, in the case of systematic linear codes) and $\bvzero$ are drawn independently using a uniform single--letter distribution $Q^{*}=\left\{\frac{1}{2},\frac{1}{2}\right\}$ (fair coin tossing).
We also define the family of the \textit{binary-input, output-symmetric (BIOS) channels}, as channels with a binary input alphabet $\calX$ ($"0"$ and $"1"$), an output alphabet $\calY$ (possibly infinite), where the transition probabilities satisfy $P(y|0)=P(-y|1), \forall y \in \calY$, for a well defined operation "$-$" (note that the definition of symmetry can be used as long as each $y \in {\cal Y}$ satisfies that $-y \in {\cal Y}$ as well). For example, the BSC, when mapping $"0"\rightarrow+1$ and $"1"\rightarrow-1$, is a BIOS channel. The additive Gaussian channel with two antipodal input letters, $x_1$ and $x_2$, is also a BIOS channel.

The following theorem is stated with regard to codes governed by the above mentioned ensembles and transmitted via a BIOS channel:

\begin{theorem}
Consider the sequence of ensembles of linear or systematic linear codes, where the elements of $\bG$ (or $\tilde{\bG}$) and $\bvzero$ are drawn independently by fair coin tossing. Let $\{P_\theta, \theta \in \Theta\}$ be a family of BIOS DMC's. Then, the lower bound $\xi_{LB}^{*}\left(R\right)$ of eq. (\ref{xi star - lower bound}), continues to hold, with $\Delta^*(P)=\ln 2 -H(P)$.
\end{theorem}
Theorem 2 is proved in Section A.1 of the Appendix.

The single--letter expression derivation for $\xi_{LB}^{*}\left(R\right)$ is carried out (see Section A.1 of the Appendix) using the same techniques as in Gallager's classical work, which are tight in the random coding sense. We therefore believe that the achievable lower bounds to the real exponential rates are tight as well.
To demonstrate the tightness of the lower bounds suggested in (\ref{xi star - lower bound}) (for general codes) and in Theorem 2 (for linear and systematic linear codes), we have the following lemma:
\begin{lemma}
Consider the family of BSC's parameterized by the crossover probability $\theta$. Then, $\xi_{LB}^{*}\left(R\right)=1$ and hence $\xi^{*}\left(R\right)=1$, in the following cases:\\
 (i) The ensemble of all codes with $Q_N(\bx)=2^{-N}$ for all $\bx$.\\
 (ii) The ensemble of linear codes and systematic linear codes, as in Theorem 2, with $\Delta^*(P)=\ln 2-H(P)$.
\end{lemma}
Lemma 1 is proved in Section V.\\
It should be mentioned that proving that under the BSC model $\xi^{*}=1$ is universally achievable by random coding over general codes and linear codes is by no means new, as it was already proved and discussed in \cite{C82}. Nevertheless, it demonstrates the tightness of $\xi_{LB}^{*}\left(R\right)$. However, to the best of our knowledge, the same result regarding ensembles of systematic linear codes has not been proved yet and is first shown here.

\subsection{Convolutional codes}

For the special case of the BSC mentioned above, we now introduce the following result, related to ensembles of time-varying convolutional codes, when the minimax decoding is used. Prior to that, we first define this ensemble and the bit error exponent related to it.

A convolutional code of rate $b/n$ ($b$, $n$ -- positive integers) and constraint length $Kb$ is defined as one for which at each time instant $t \geq 0$, the code vector of length $n$, $\bvt$, is obtained by 
\begin{equation}
\label{convolutional code definition}
\bvt=\sum_{j=0}^{\min\{t,K-1\}} \butminusj \bGj \oplus \bvzero,
\end{equation}
where $\butminusj$ is a binary information row vector of length $b$ at time $t-j$, $\bGj, 0 \leq j \leq K-1$, are binary matrices with $b$ rows and $n$ columns each, and $\bvzero$ is a vector of length $n$.

Let us now consider a code ${\cal C}$, governed by i.i.d. random coding over the ensemble of time-varying convolutional codes, whose code vector of time instant $t \geq 0$, $\bvt$, is obtained by
\begin{equation}
\label{ensemble of time-varying convolutional codes definition}
\bvt=\sum_{j=0}^{\min\{t,K-1\}} \butminusj \bGjt\oplus\bvzerot,
\end{equation}
where at each time instant $t$, the elements of $\bGjt, 0 \leq j \leq K-1$ and $\bvzerot$ are drawn independently using the uniform single--letter distribution $\left\{\frac{1}{2},\frac{1}{2}\right\}$.

The average bit error probability, $\overline{P_b(\Omega_K)}$, associated with a sequence of decoders $\Omega_K=\left\{\Omega_{K,N}\right\}_{N=1}^{\infty}$ of block length $N$ and constraint length $K$, and averaged over the ensemble of time-varying convolutional codes, is defined as the expected relative frequency of bit errors in the decoded information stream, i.e. 
\begin{equation}
\label{bit error probability}
\overline{P_b(\Omega_K)}=\limsup_{N\to\infty} \overline{P_b(\Omega_{K,N})}.
\end{equation}
The bit error exponent associated with a sequence of decoders $\Omega = \left\{\Omega_K\right\}_{K=1}^{\infty}$ is defined as 
\begin{equation}
\label{bit error exponent}
\overline{E_b(\Omega)}= - \limsup_{K\to\infty} \frac{1}{K}\ln\overline{P_b(\Omega_K)}.
\end{equation}

\begin{theorem}
Consider the sequence of ensembles of time--varying convolutional codes of rate $b/n$ and constraint length $Kb$ (with $K\rightarrow \infty$), described as in the previous paragraph, and assume a family of BSC's parameterized by the crossover probability $\theta$.\\
The achievable bit error exponent (as defined in (\ref{bit error exponent})) using the minimax decoder is equal to the one when $\theta$ is known and the ML decoder is used.
\end{theorem}
The proof of this theorem is based on the following observation:\\
Under the BSC model with an unknown crossover probability $\theta$, the minimax decision rule (as defined in (\ref{explicit_decision_rule_2})) is equivalent to a decision rule, denoted by $\Lambda$, and defined as:
\begin{equation}
\label{rho based decision rule}
\Lambda_{m} =
	\left\{
	\by |
			\rho(\bxm,\by) 
			\leq
			\rho(\bxmtag,\by),\ \forall{m'}\neq m
	\right\},
\end{equation}
with ties broken arbitrarily,
where 
\begin{equation}
\label{rho definition}
\rho(\bx,\by)
=
\min\left\{
					\delta{(\bx,\by)} ,
					1-\delta{(\bx,\by)}
		\right\}.
\end{equation}
As mentioned in Section II, $\delta(\bx,\by)$ denotes the normalized Hamming distance between $\bx$ and $\by$. 
This equivalence is proved in Section A.7 of the Appendix. We should note that for this case, the minimax decoder coincides with the MMI decoder as well. Based on this equivalence, the full proof of Theorem 3 is given in Section VII. We also introduce an efficient implementation of minimax decoding, based on a slightly modified version of the Viterbi algorithm. This is done by applying the Viterbi algorithm twice: first for minimum
Hamming distance, and then for maximum Hamming distance.
This process results in two survivors and the selection between them is done in favor of the one whose normalized Hamming metric is more distant from $\frac{1}{2}$ (the one with the minimal $\rho$).

\section{Proof of Theorem 1}

We first observe that for a DMC, $\left\{P_\theta \left(y|x \right),x\in {\cal X},y\in {\cal Y}\right\}$, and for each vector pair 
$
(\bx,\by)
$, the minimax metric for a given $\theta$, 
$
f_\theta(\bx,\by)
$, 
depends on $\bx$ and $\by$ only via their joint empirical distribution:
\begin{eqnarray}
%
\label{f dep on x n y-5}
f_\theta(\bx,\by)
=
			\hat{E}_{\bx\by}
				\ln{
						P_\theta({Y}|{X})
					}
			+{\xi E_{r}^{*}(\theta)}.
\end{eqnarray}

We, therefore, conclude that the value of $\theta$ maximizing $f_{\theta}(\bx,\by)$ also depends on $\bx$ and $\by$ only via their joint empirical distribution.
Let $\Theta_N$ denote the subset of $\Theta$ with values of $\theta$ that achieve 
$
\max_\theta	f_{\theta}(\bx,\by) = f(\bx,\by)
$
as $(\bx,\by)$ exhaust ${{\cal X}^N}\times{{\cal Y}^N}$. In the decoding process, maximization over $\theta$ can be achieved only by points in $\Theta_N$.  Since the number of joint empirical distributions of 
$
(\bx,\by)
$
is upper bounded by $\left(N+1\right)^{\left|X\right|\left|Y\right|}$, then
$\left|\Theta_N\right| \leq \left(N+1\right)^{\left|X\right|\left|Y\right|}$
as well.

As a first step, we assume given channel input and output vectors, $\bx$ and $\by$, respectively. Considering a random coding distribution, $Q_N$, we exponentially evaluate the probability of having another codeword $\bxtag$ that is preferred by the minimax decoder over $\bx$. This probability will be denoted by $a(\bx,\by)$.
\begin{eqnarray}
a(\bx,\by)
& = &
Q_N\left\{
	f(\bXtag,\by) \geq f(\bx,\by)
	\right\}
\nonumber
\\
& = &
Q_N\left\{
	\max_{\theta' \in \Theta_N}f_{\theta'}(\bXtag,\by) \geq f(\bx,\by)
	\right\}	
\nonumber
\\
& \stackrel{\stackrel{(a)}{\cdot}}{=} &
\max_{\theta' \in \Theta_N}
Q_N\left\{
	f_{\theta'}(\bXtag,\by) \geq f(\bx,\by)
	\right\}	
\nonumber
\\
& = &
\max_{\theta' \in \Theta_N}
Q_N\left\{
	\sum_{i=1}^{N}
	\ln P_{\theta'}(y_i|X'_i)
	\geq 
	-N \xi E_r^{*}(\theta')+N\cdot f(\bx,\by)
	\right\}		
\nonumber
\\
& \stackrel{\stackrel{(b)}{\cdot}}{=} &
\max_{\theta' \in \Theta_N}
\min_{s \geq 0}
E_{Q_N}
\Biggl[
	\exp\biggl\{
		s
			\bigl[
				\sum_{i=1}^{N}
				\ln P_{\theta'}(y_i|X'_i)
				+N \xi E_r^{*}(\theta')
			\bigr]
		\biggr\}
\Biggr]
\cdot					
	\exp\left\{
		-sN f(\bx,\by)
		\right\}
\nonumber
\\
& = &
\max_{\theta' \in \Theta_N}
\min_{s \geq 0}
E_{Q_N}
	e^{
			sN f_{\theta'}(\bXtag,\by)
		}
\cdot					
	e^{
		-sN f(\bx,\by)
		},
\end{eqnarray}
where ($a$) is true since
\begin{eqnarray}
\max_{\theta' \in \Theta_N}
Q_N\left\{
	f_{\theta'}(\bXtag,\by) \geq f(\bx,\by)
	\right\}	
& \leq &
Q_N\left\{
	\max_{\theta' \in \Theta_N}f_{\theta'}(\bXtag,\by) \geq f(\bx,\by)
	\right\}
\nonumber
\\
& = &
Q_N\left\{
	\bigcup_{\theta' \in \Theta_N}f_{\theta'}(\bXtag,\by) \geq f(\bx,\by)
	\right\}
\nonumber
\\
& \leq &
\sum_{\theta' \in \Theta_N}
Q_N\left\{
	f_{\theta'}(\bXtag,\by) \geq f(\bx,\by)
\right\}
\nonumber
\\
& = &
\left|\Theta_N\right| \cdot
\max_{\theta' \in \Theta_N}
Q_N\left\{
	f_{\theta'}(\bXtag,\by) \geq f(\bx,\by)
	\right\},	
\end{eqnarray}
and in ($b$) we used the Cheroff bound, which is tight in the exponential sense.

By using the method of types, it is proved in Section A.3 of the Appendix that for any real $\alpha$,
\begin{eqnarray}
\label{E upper bound}
E_{Q_N}
		\bigl[
					{e^{N{\alpha}{f_{\theta}(\bX,\by)}
						}}
		\bigr]
\stackrel{\cdot}{=}
e^{
	N\bigl[ 
			 	{\alpha}{\xi E_{r}^{*}(\theta)}
			 	-
				\min_{P_{\bx|\by}}
				 	A(\theta,\alpha,P_{\bx\by})
   \bigr]
	},
\end{eqnarray}
where the function $A(\theta,\alpha,P_{xy})$ is defined as in (\ref{A_theta}).\\
Using this observation, we can continue to evaluate $a(\bx,\by)$ as follows:
\begin{eqnarray}
a(\bx,\by)
& \stackrel{\cdot}{=} &
\max_{\theta' \in \Theta_N}
\min_{s \geq 0}
\exp\Bigl\{
		N
			\bigl[
				s \xi E_r^{*}(\theta')
				-\min_{P_{\bxtag | \by}} A(\theta',s,P_{\bxtag \by})
			\bigr]
	\Bigr\}
\cdot					
	\exp\left\{
		-sN f(\bx,\by)
		\right\}
\nonumber
\\
& = &
\max_{\theta' \in \Theta_N}
\min_{s \geq 0}
\exp\Bigl\{
		-N
			\bigl[
				-s \xi E_r^{*}(\theta')
				+\min_{P_{\bxtag | \by}}A(\theta',s,P_{\bxtag \by})
				+s f(\bx,\by)
			\bigr]
		\Bigr\}		
\nonumber
\\
& \stackrel{\Delta}{=} &
\max_{\theta' \in \Theta_N}
\min_{s \geq 0}
\exp\Bigl\{
		-N
			\bigl[
				G(\theta',s,\xi,P_{\bx \by})
			\bigr]
		\Bigr\}.
\end{eqnarray}
Therefore, the probability that the decoder will prefer any of the other $M-1$ codevectors rather than the transmitted codevector $\bx$ can be evaluated as follows:
\begin{eqnarray}
1-(1-a(\bx,\by))^{M-1}
& \stackrel{\stackrel{(a)}{\cdot}}{=} &
\min\{1,e^{NR}\cdot a(\bx,\by)\}
\nonumber
\\
& \stackrel{\cdot}{=} &
\min\left\{
	1,
	\max_{\theta' \in \Theta_N}
			\min_{s \geq 0}
			\exp\left\{
						-N
						\bigl[
							G(\theta',s,\xi,P_{\bx \by})-R
						\bigr]
					\right\}
\right\}
\nonumber
\\
& = &
\max_{\theta' \in \Theta_N}
\min_{s \geq 0}
\exp\Bigl\{
		-N
		 \cdot
			\max_{0 \leq \rho \leq 1}
			\rho
			\bigl[
				G(\theta',s,\xi,P_{\bx \by})-R
			\bigr]
		\Bigr\}		
\nonumber
\\
& = &
\max_{\theta' \in \Theta_N}
\min_{\stackrel{0 \leq \rho \leq 1}{s \geq 0}}
\exp\Bigl\{
		-N
		 \cdot
			\bigl[
			\rho	G(\theta',s,\xi,P_{\bx \by})
			-\rho R
			\bigr]
		\Bigr\},			
\end{eqnarray}
where the equivalence in ($a$) (see \cite{SBM07}, Section V, and \cite{S03}, Section A.2 p. 109-110) implies that the union bound in the random coding error exponent is tight. 

Now, we will evaluate $\overline{S}_N$, the average of the minimax criterion over the ensemble of codebooks governed by a random coding distribution, for the minimax decoder defined in (\ref{explicit_decision_rule_2}):
\begin{eqnarray}
\overline{S}_N
& \stackrel{\cdot}{=} &
\max_{\theta \in \Theta}
	\left\{
		\frac
			{\overline{P}_{E}\left(\Omega|\theta\right)}
			{e^{-N\xi E_{r}^{*}(\theta)}}
	\right\}
\nonumber
\\
& = &
\max_{\theta \in \Theta}
	\left\{
			e^{N\xi E_{r}^{*}(\theta)}
			\sum_{\bx \in {\cal X}^N}	
			Q_N(\bx)
			\sum_{\by \in {\cal Y}^N}
			P_{\theta}(\by|\bx)
			\left[
				1-(1-a(\bx,\by))^{M-1}
			\right]
	\right\}
\nonumber
\\
& \stackrel{\cdot}{=} &
\max_{\theta \in \Theta}
	\left\{
			\sum_{\bx \in {\cal X}^N}
			Q_N(\bx)
			\sum_{\by \in {\cal Y}^N}
			e^{N f_{\theta}(\bx,\by)}
			\max_{\theta' \in \Theta_N}
			\min_{\stackrel{0 \leq \rho \leq 1}{s \geq 0}}
			\exp\Bigl\{
					-N
						\bigl[
							\rho\	G(\theta',s,\xi,P_{\bx \by})
							-\rho R
						\bigr]
					\Bigr\}
	\right\}	
\nonumber
\\
& \stackrel{\stackrel{(a)}{\cdot}}{=} &
\max_{\theta \in \Theta}
	\Biggl\{
			\sum_{T_{\bx \by} \subset {\cal X}^N \times {\cal Y}^N}
			Q_N(T_{\bx})
			\left|T_{\by | \bx}\right|
			e^{N f_{\theta}(\bx,\by)}
			\max_{\theta' \in \Theta_N}
			\min_{\stackrel{0 \leq \rho \leq 1}{s \geq 0}}
\nonumber\\
& &		
\ \ \ \ \ \ \ \ \	
				e^{N \rho s \xi E_r^{*}(\theta')}
				\cdot
				e^{-N \rho \min_{P_{\bxtag | \by}} A(\theta',s,P_{\bxtag \by})}
				\cdot
				e^{-N \rho s f(\bx,\by)}
				\cdot
				e^{N \rho R}		 			
	\Biggr\}		
\nonumber
\\
& \stackrel{\cdot}{=} &
\max_{P_{\bx \by}}
	\Biggl\{
			e^{-N\Delta^{*}_N(P_{\bx})}
			\cdot
			e^{N H_{\bx\by}(Y|X)}
			\left[
				\max_{\theta \in \Theta_N}			
				e^{N f_{\theta}(\bx,\by)}
			\right]
			\max_{\theta' \in \Theta_N}
			\min_{\stackrel{0 \leq \rho \leq 1}{s \geq 0}}
			\max_{P_{\bxtag | \by}}			
\nonumber\\
& &		
\ \ \ \ \ \ \ \ \	
				e^{N \rho s \xi E_r^{*}(\theta')}
				\cdot
				e^{-N \rho A(\theta',s,P_{\bxtag \by})}
				\cdot
				\left[
					\max_{\theta'' \in \Theta_N}			
					e^{N f_{\theta''}(\bx,\by)}
				\right]^{-\rho s}
				\cdot
				e^{N \rho R}		 			
	\Biggr\}	
\nonumber
\\
& \stackrel{\stackrel{(b)}{\cdot}}{=} &
\max_{P_{\bx \by}}
\max_{\theta' \in \Theta_N}
\min_{\stackrel{0 \leq \rho \leq 1}{s \geq 0}}
\max_{P_{\bxtag | \by}}
	\Biggl\{
			e^{-N \Delta^{*}(P_{\bx})}
			e^{N H_{\bx\by}(Y|X)}
			e^{N \rho s \xi E_r^{*}(\theta')}
			e^{-N \rho A(\theta',s,P_{\bxtag \by})}
\nonumber\\
& &
\ \ \ \ \ \ \ \ \	\ \ \ \ \ \ \ \ \	\ \ \ \ \ \ \ \	\ \ \ \ \ 
			\left[
				\max_{\theta \in \Theta_N}			
				e^{N f_{\theta}(\bx,\by)}
			\right]^{1-\rho s}
				e^{N \rho R}		 			
	\Biggr\}		
\nonumber
\\
& = &
\max_{P_{\bx \by}}
\max_{\theta' \in \Theta_N}
\min_{\stackrel{0 \leq \rho \leq 1}{s \geq 0}}
\max_{P_{\bxtag | \by}}
	\Biggl\{
			\exp
				\biggl\{
						N
						\bigl[
							-\Delta^{*}(P_{\bx})
							+H_{\bx\by}(Y|X)
							+\rho s \xi E_r^{*}(\theta')
\nonumber\\
& &
\ \ \ \ \ \ \ \ \	\ \ \ \ \ \ \ \ \	\ \ \ \ \ \ \ \	\ \ 
							-\rho A(\theta',s,P_{\bxtag \by})
							+\rho R
					\bigr]							
				\biggr\}
			\left[
				\max_{\theta \in \Theta_N}			
				e^{N f_{\theta}(\bx,\by)}
			\right]^{1-\rho s}
	\Biggr\}		
\nonumber
\\
& \stackrel{\Delta}{=} &
\max_{P_{\bx \by}}
\max_{\theta' \in \Theta_N}
\min_{\stackrel{0 \leq \rho \leq 1}{s \geq 0}}
\max_{P_{\bxtag | \by}}
	\Biggl\{
			\exp\left\{
						N
						\cdot
						T(\theta',P_{\by},P_{\bx|\by},P_{\bxtag|\by},s,\rho,\xi,R )
				\right\}
\nonumber\\
& &
\ \ \ \ \ \ \ \ \	\ \ \ \ \ \ \ \ \	\ \ \ \ \ \ \ \	\ \ 
			\left[
				\max_{\theta \in \Theta_N}			
				e^{N f_{\theta}(\bx,\by)}
			\right]^{1-\rho s}
	\Biggr\},
\end{eqnarray}
where in ($a$) we switched to a summation over the joint empirical types of $\bx$ and $\by$ (which is legitimate since both $f_{\theta}(\bx,\by)$ and $G(\theta',s,\xi,P_{\bx \by})$ depend on $\bx$ and $\by$ via their joint empirical distribution), and in ($b$), we used the convergence assumption of the random coding distributions within the class $\calQ$ to claim that $\Delta^{*}_N(P_{\bx}) \rightarrow \Delta^{*}(P_{\bx})$ as $N\rightarrow  \infty$ independently of $P_{\bx}$, and also united the optimizations over $\theta$ and $\theta''$.

We should observe that:

\begin{lefteqn}
{
\min_{\stackrel{0 \leq \rho \leq 1}{s \geq 0}}
\max_{P_{\bxtag | \by}}
	\left\{
			\exp\left\{
						N
						\cdot
						T(\theta',P_{\by},P_{\bx|\by},P_{\bxtag|\by},s,\rho,\xi,R )
				\right\}
			\left[
				\max_{\theta \in \Theta_N}			
				e^{N f_{\theta}(\bx,\by)}
			\right]^{1-\rho s}
	\right\}		
 = 
}
\end{lefteqn}
\begin{eqnarray}
& = &
\min 
\Biggl\{
\min_{\stackrel{0 \leq \rho \leq 1}{0 \leq s \leq 1/\rho}}
\max_{P_{\bxtag | \by}}
\max_{\theta \in \Theta_N}		
\exp
\biggl\{
N
\left[
T(\theta',P_{\by},P_{\bx|\by},P_{\bxtag|\by},s,\rho,\xi,R )
+
f_{\theta}(\bx,\by)(1-\rho s)
\right]
\biggr\}
,
\nonumber\\
&  &\ \ \ \ \ \ \ \ \ 
\min_{\stackrel{0 \leq \rho \leq 1}{s \geq  1/\rho}}
\max_{P_{\bxtag | \by}}
\min_{\theta \in \Theta_N}		
\exp
\biggl\{
N
\left[
T(\theta',P_{\by},P_{\bx|\by},P_{\bxtag|\by},s,\rho,\xi,R )
+
f_{\theta}(\bx,\by)(1-\rho s)
\right]
\biggr\}
\Biggr\}
\nonumber\\
& \stackrel{(a)}{=} &
\label{inner_term_1_optimizations_interchange-1}
\min 
\Biggl\{
\max_{\theta \in \Theta_N}
\min_{\stackrel{0 \leq \rho \leq 1}{0 \leq s \leq 1/\rho}}
\max_{P_{\bxtag | \by}}
\exp
\biggl\{
N
\left[
T(\theta',P_{\by},P_{\bx|\by},P_{\bxtag|\by},s,\rho,\xi,R )
+
f_{\theta}(\bx,\by)(1-\rho s)
\right]
\biggr\}
,
\nonumber\\
\label{inner_term_1_optimizations_interchange-2}
&  &\ \ \ \ \ \ \ \ \ 
\min_{\theta \in \Theta_N}		
\min_{\stackrel{0 \leq \rho \leq 1}{s \geq  1/\rho}}
\max_{P_{\bxtag | \by}}
\exp
\biggl\{
N
\left[
T(\theta',P_{\by},P_{\bx|\by},P_{\bxtag|\by},s,\rho,\xi,R )
+
f_{\theta}(\bx,\by)(1-\rho s)
\right]
\biggr\}
\Biggr\}
\nonumber\\
\label{inner_term_1_optimizations_interchange-3}
\\
& \stackrel{\Delta}{=} &
\min 
\Biggl\{
\max_{\theta \in \Theta_N}		
\min_{\stackrel{0 \leq \rho \leq 1}{0 \leq s \leq 1/\rho}}
\max_{P_{\bxtag | \by}}
\exp\left\{
	N
	\cdot
	\tilde{T}(\theta,\theta',P_{\by},P_{\bx|\by},P_{\bxtag|\by},s,\rho,\xi,R )
\right\}
,
\nonumber\\
&  & \ \ \ \ \ \ \ 
\min_{\theta \in \Theta_N}		
\min_{\stackrel{0 \leq \rho \leq 1}{s \geq  1/\rho}}
\max_{P_{\bxtag | \by}}
\exp\left\{
	N
	\cdot
	\tilde{T}(\theta,\theta',P_{\by},P_{\bx|\by},P_{\bxtag|\by},s,\rho,\xi,R )
\right\}
\Biggr\},
\end{eqnarray}
where in ($a$), two interchanges are made: one between the minimization over $\rho$ and $s$ and the maximization over $\theta$ in the left term of the outer minimization, and one between the maximization over $P_{\bxtag|\by}$ and the minimization over $\theta$ in the right term of the outer minimization. The first interchange is justified in the Appendix, Section A.2. The second interchange is possible since the term to be optimized is a product of two exponential terms, one depends on $P_{\bxtag|\by}$ and one depends on $\theta$, therefore the optimizations can be done independently.

Consequently, we conclude that:
\begin{eqnarray}
\overline{S}_N
& \stackrel{\cdot}{=} &
	\max_{P_{\bx \by}}
	\max_{\theta' \in \Theta_N}
  \min
  \Biggl\{
	  \max_{\theta \in \Theta_N}		
	  \min_{\stackrel{0 \leq \rho \leq 1}{0 \leq s \leq 1/\rho}}
		\max_{P_{\bxtag | \by}}
		\exp
		\left\{
			N
			\cdot
			\tilde{T}(\theta,\theta',P_{\by},P_{\bx|\by},P_{\bxtag|\by},s,\rho,\xi,R )
		\right\}
		,
		\nonumber\\
&  & \ \ \ \ \ \ \ \ \ \ \ \ \ \ \ \ \ \ \ \ \ \ 
		\min_{\theta \in \Theta_N}		
		\min_{\stackrel{0 \leq \rho \leq 1}{s \geq  1/\rho}}
		\max_{P_{\bxtag | \by}}
		\exp
		\left\{
			N
			\cdot
			\tilde{T}(\theta,\theta',P_{\by},P_{\bx|\by},P_{\bxtag|\by},s,\rho,\xi,R )
		\right\}
	\Biggr\}
\nonumber\\
& = &
	\max_{P_{\bx \by}}
	\max_{\theta' \in \Theta_N}
  \min
  \Biggl\{
		\exp
		\biggl\{
		N
			\cdot
		\max_{\theta \in \Theta_N}
		\min_{\stackrel{0 \leq \rho \leq 1}{0 \leq s \leq 1/\rho}}
		\max_{P_{\bxtag | \by}}
			\tilde{T}(\theta,\theta',P_{\by},P_{\bx|\by},P_{\bxtag|\by},s,\rho,\xi,R )
		\biggr\}
		,
		\nonumber\\
&  & \ \ \ \ \ \ \ \ \ \ \ \ \ \ \ \ \ \ \ \ \ \ 
		\exp
		\biggl\{
		N
		\cdot
			\min_{\theta \in \Theta_N}		
			\min_{\stackrel{0 \leq \rho \leq 1}{s \geq  1/\rho}}
			\max_{P_{\bxtag | \by}}
			\tilde{T}(\theta,\theta',P_{\by},P_{\bx|\by},P_{\bxtag|\by},s,\rho,\xi,R )
		\biggr\}
	\Biggr\}
\nonumber\\
& = &
\exp
\Biggl\{
N
\cdot
	\max_{P_{\bx \by}}
	\max_{\theta' \in \Theta_N}
  \min
		\biggl\{
			\max_{\theta \in \Theta_N}
			\min_{\stackrel{0 \leq \rho \leq 1}{0 \leq s \leq 1/\rho}}
			\max_{P_{\bxtag | \by}}
			\tilde{T}(\theta,\theta',P_{\by},P_{\bx|\by},P_{\bxtag|\by},s,\rho,\xi,R )
		,
\nonumber\\
&  & \ \ \ \ \ \ \ \ \ \ \ \ \ \ \ \ \ \ \ \ \ \ \ \ \ \ \ \ \ \ \ \ \ 
			\min_{\theta \in \Theta_N}		
			\min_{\stackrel{0 \leq \rho \leq 1}{s \geq  1/\rho}}
			\max_{P_{\bxtag | \by}}
			\tilde{T}(\theta,\theta',P_{\by},P_{\bx|\by},P_{\bxtag|\by},s,\rho,\xi,R )
		\biggr\}
\Biggr\}
.
\nonumber\\
\end{eqnarray}		
Now,

\begin{lefteqn}
{
\tilde{T}(\theta,\theta',P_{\by},P_{\bx|\by},P_{\bxtag|\by},s,\rho,\xi,R ) =
}
\end{lefteqn}
\begin{eqnarray}
& = &
			-\Delta^{*}(P_{\bx})
			+H_{\by}(Y)
			-I_{\bx\by}(X;Y)
			+ \rho s \xi E_r^{*}(\theta')
			-\rho A(\theta',s,P_{\bxtag \by})
\nonumber\\
& &
			+(1-\rho s)\hat{E}_{\bx\by}
										\ln{
												P_\theta(Y|X)
												}
			+(1-\rho s)\xi E_r^{*}(\theta)
			+\rho R
\nonumber\\
& = &
			-A(\theta,1-\rho s,P_{\bx \by})
			-\rho A(\theta',s,P_{\bxtag \by})
			+H_{\by}(Y)
			+\rho s \xi E_r^{*}(\theta')
\nonumber\\
& &
			+(1-\rho s)\xi E_r^{*}(\theta)
			+\rho R
\nonumber
\\
& = &
			-B(\theta,\theta',P_{\by},P_{\bx| \by},P_{\bxtag | \by},s,\rho)
			+\rho s \xi E_r^{*}(\theta')
			+(1-\rho s)\xi E_r^{*}(\theta)
			+\rho R,
\nonumber
\\			
\end{eqnarray}
where the function $B(\theta,\theta',P_{\by},P_{\bx| \by},P_{\bxtag | \by},s,\rho)$ is defined as in (\ref{B_definition}).

Therefore, in order for $\overline{S}_N$ to grow sub--exponentially with $N$, we seek the maximal $\xi$ such that:
\begin{eqnarray}
	\max_{P_{\bx \by}}
	\max_{\theta' \in \Theta_N}
&  \min &
		\biggl\{
			\max_{\theta \in \Theta_N}
			\min_{\stackrel{0 \leq \rho \leq 1}{0 \leq s \leq 1/\rho}}
			\max_{P_{\bxtag | \by}}
			\tilde{T}(\theta,\theta',P_{\by},P_{\bx|\by},P_{\bxtag|\by},s,\rho,\xi,R )
		,
\nonumber\\
& &
\ \ \ \ \ 
			\min_{\theta \in \Theta_N}		
			\min_{\stackrel{0 \leq \rho \leq 1}{s \geq  1/\rho}}
			\max_{P_{\bxtag | \by}}
			\tilde{T}(\theta,\theta',P_{\by},P_{\bx|\by},P_{\bxtag|\by},s,\rho,\xi,R )
		\biggr\}
\leq 0
\end{eqnarray}
As the empirical distributions become dense in continuum of probability distributions as $N\rightarrow\infty$, and since the function $\tilde{T}(\theta,\theta',P_{\by},P_{\bx|\by},P_{\bxtag|\by},s,\rho,\xi,R )$ is continuous in $P_{\by}$, $P_{\bx|\by}$ and $P_{\bxtag|\by}$, it is equivalent to perform the above optimizations over continuous distributions rather than empirical distributions. The same token can be used in order to broaden the maximization space for $\theta$ and $\theta'$ from $\Theta_N$ to $\Theta$. Thus, the condition becomes:
\begin{eqnarray}
	\max_{P_{X y}}
	\max_{\theta' \in \Theta}
&  \min &
		\biggl\{
			\max_{\theta \in \Theta}
			\min_{\stackrel{0 \leq \rho \leq 1}{0 \leq s \leq 1/\rho}}
			\max_{P_{X' | y}}
			\tilde{T}(\theta,\theta',P_{y},P_{X|y},P_{X'|y},s,\rho,\xi,R )
		,
\nonumber\\
& &
\ \ \ \ \ 
			\min_{\theta \in \Theta}		
			\min_{\stackrel{0 \leq \rho \leq 1}{s \geq  1/\rho}}
			\max_{P_{X' | y}}
			\tilde{T}(\theta,\theta',P_{y},P_{X|y},P_{X'|y},s,\rho,\xi,R )
		\biggr\}
\leq 0
\end{eqnarray}

In other words, a maximal $\xi$ is sought such that:

\begin{lefteqn}
{	\forall{P_{X y}},
	\forall{\theta' \in \Theta}
}
\end{lefteqn}
\begin{eqnarray}
{			\max_{\theta \in \Theta}
			\min_{\stackrel{0 \leq \rho \leq 1}{0 \leq s \leq 1/\rho}}
			\max_{P_{X' | y}}
			\tilde{T}(\theta,\theta',P_{y},P_{X|y},P_{X'|y},s,\rho,\xi,R )
			\leq 0
}
\end{eqnarray}
or
\begin{eqnarray}
{			\min_{\theta \in \Theta}		
			\min_{\stackrel{0 \leq \rho \leq 1}{s \geq  1/\rho}}
			\max_{P_{X' | y}}
			\tilde{T}(\theta,\theta',P_{y},P_{X|y},P_{X'|y},s,\rho,\xi,R )
			\leq 0
}
\end{eqnarray}			
An equivalent condition is:

\begin{lefteqn}
{
	\forall{P_{X y}},
	\forall{\theta' \in \Theta}	
	\nonumber
}
\end{lefteqn}	
\begin{eqnarray}
	\xi \leq
	\min_{\theta \in \Theta}			
	\max_{\stackrel{0 \leq \rho \leq 1}{0 \leq s \leq 1/\rho}}
	\min_{P_{X' | y}}
	\frac{B(\theta,\theta',P_{y},P_{X| y},P_{X' | y},s,\rho)-\rho R}
			{(1-\rho s) E_r^{*}(\theta)+\rho s E_r^{*}(\theta')}
\end{eqnarray}
 	 or
\begin{eqnarray}
	\xi \leq
	\max_{\theta \in \Theta}
	\max_{\stackrel{0 \leq \rho \leq 1}{s  \geq 1/\rho}}
	\min_{P_{X' | y}}
	\frac{B(\theta,\theta',P_{y},P_{X| y},P_{X' | y},s,\rho)-\rho R}
			{(1-\rho s) E_r^{*}(\theta)+\rho s E_r^{*}(\theta')}
\end{eqnarray}
Therefore,
\begin{eqnarray}
\xi^{*}(R)
& = &
\min_{P_{X y}}
\min_{\theta' \in \Theta}
\max
\Bigl\{
\min_{\theta \in \Theta}
\max_{\stackrel{0 \leq \rho \leq 1}{0 \leq s \leq 1/\rho}}
\min_{P_{X' | y}}
\frac{B(\theta,\theta',P_{y},P_{X| y},P_{X' | y},s,\rho)-\rho R}
			{(1-\rho s) E_r^{*}(\theta)+\rho s E_r^{*}(\theta')}
,
\nonumber\\
& &
\ \ \ \ \ \ \ \ \ \ \ \ \ \ \ \ \ \ \ \ \ 
\max_{\theta \in \Theta}
\max_{\stackrel{0 \leq \rho \leq 1}{s  \geq 1/\rho}}
\min_{P_{X' | y}}
\frac{B(\theta,\theta',P_{y},P_{X| y},P_{X' | y},s,\rho)-\rho R}
			{(1-\rho s) E_r^{*}(\theta)+\rho s E_r^{*}(\theta')}
\Bigr\}.
\end{eqnarray}

\section{Example - the BSC}

In this section, we demonstrate that for the special case of BSC with an unknown crossover probability, and a uniform random coding distribution, $\xi_{LB}^{*}(R)=1$ and hence $\xi^{*}(R)=1$, in agreement with well known results \cite{C82}.


Consider the lower bound (\ref{xi star - lower bound}) and choose the uniform single--letter random coding distribution $Q^{*}=\{\frac{1}{2},\frac{1}{2}\}$.\\
Now, the value of $A(\theta,\alpha,P_{X Y})$ is (see (\ref{A_theta_for i.i.d.})):
\begin{eqnarray}
\label{Tightness of the LB - A - 2}
A(\theta,\alpha,P_{X Y})
& = &
								\ln 2
						 		-H(X|Y)
						 		-{\alpha}E
												\ln{
													P_\theta(Y|X)
													}
\end{eqnarray}
Therefore, 
\begin{eqnarray}
\label{Tightness of the LB - A - 6}
	\min_{P_{X|Y}}
			A(\theta,\alpha,P_{X Y})
& = &
\ln 2						 
-\max_{P_{X|Y}}
								\left\{
 									H(X|Y)
						 			+{\alpha}E
												\ln{
													P_\theta(Y|X)
												}
								\right\}
\end{eqnarray}
In addition, for the case of BSC with an unknown crossover probability, $\theta$, we have (see \cite{UD_WE}, Section VI):
\begin{eqnarray}
\label{BSC - maximization upper bound - 1}
\max_{P_{X|Y}}
	\Bigl\{
		H(X|Y)
		+{\alpha}E
					\ln{
						P_\theta(Y|X)
					}
	\Bigr\}
& = &
\ln\bigl[
			\left(
				1-\theta
			\right)
			^\alpha
			+\theta^\alpha
		\bigr]
\nonumber\\		
& \stackrel{\Delta}{=} &
\calV(\theta,\alpha)
\label{BSC - maximization upper bound - 2}
\end{eqnarray}
From these two observations, we conclude that:
\begin{eqnarray}
\label{Tightness of the LB - 3}
	\min_{P_{X|Y}}
			A(\theta,\alpha,P_{X Y})
& =	&
\ln2-
\calV(\theta,\alpha)
\end{eqnarray}
Using (\ref{B_definition}), we get:
\begin{eqnarray}
\label{BSC - xi for Q-star - 2}
\xi_{LB}^{*}\left(R\right)
& = &
\min_{P_{Y}}
\min_{\theta,\theta' \in \Theta}
\max_{\stackrel{0\leq\rho\leq1}{0\leq\lambda\leq 1/\rho}}
\min_{P_{X|Y}}
\min_{P_{X'|Y}}
		\frac{
				A(\theta,1-\lambda \rho,P_{X Y})
				+\rho
 				A(\theta',\lambda,P_{X' Y})
			-H(Y)
			-\rho R
		}
		{
			(1- \lambda \rho) \cdot E_{r}^{*}(\theta)
			+\lambda \rho \cdot E_{r}^{*}(\theta')
		}
\nonumber\\
\label{BSC - xi for Q-star - 2.5}
& = &
\min_{P_{Y}}
\min_{\theta,\theta' \in \Theta}
\max_{\stackrel{0\leq\rho\leq1}{0\leq\lambda\leq 1/\rho}}
		\frac{
				(1+\rho)\ln2
				-\calV(\theta,1- \lambda \rho)
				-\rho \calV(\theta',\lambda)
				-H(Y)
				-\rho R
		}
		{
			(1- \lambda \rho) \cdot E_{r}^{*}(\theta)
			+\lambda \rho \cdot E_{r}^{*}(\theta')
		}
\nonumber\\
\label{BSC - xi for Q-star - 3}
& \geq &
\min_{\theta,\theta' \in \Theta}
\max_{\stackrel{0\leq\rho\leq1}{0\leq\lambda\leq 1/\rho}}
\min_{P_{Y}}
		\frac{
				(1+\rho)\ln2
				-\calV(\theta,1- \lambda \rho)
				-\rho \calV(\theta',\lambda)
				-H(Y)
				-\rho R
		}
		{
			(1- \lambda \rho) \cdot E_{r}^{*}(\theta)
			+\lambda \rho \cdot E_{r}^{*}(\theta')
		}
\nonumber\\
\label{BSC - xi for Q-star - 4}
& = &
\min_{\theta,\theta' \in \Theta}
\max_{\stackrel{0\leq\rho\leq1}{0\leq\lambda\leq 1/\rho}}
		\frac{
				\rho \ln2
				-\calV(\theta,1- \lambda \rho)
				-\rho \calV(\theta',\lambda)
				-\rho R
		}
		{
			(1- \lambda \rho) \cdot E_{r}^{*}(\theta)
			+\lambda \rho \cdot E_{r}^{*}(\theta')
		}.		
\end{eqnarray}
Now, the random coding error exponent associated with ML decoding, $E_{r}^{*}(\theta)$, to which the minimax decoding error exponent is compared, is achieved for the BSC model by the following optimization (see \cite[Sect. 3.1, 3.2 and 3.4]{VO79}):
\begin{eqnarray}
\label{ML error_exponent - 1}
E_{r}^{*}(\theta)
& = &
\max_{0\leq\rho\leq1}
\max_{Q}
\biggl\{
	-\ln
	\sum\limits_{y \in \{0,1\}}
	\Bigl[
		\sum\limits_{x \in \{0,1\}}
			Q(x)
			\cdot
			P_\theta \left(y|x \right)
			^{\frac{1}{1+\rho}}
	\Bigr]
	^{1+\rho}
	-\rho R
\biggr\}
\nonumber\\
\label{ML error_exponent - 2}
& \stackrel{(a)}{=} & 
\max_{0\leq\rho\leq1}
\Bigl\{
			\rho \ln2
			-\left(1+\rho\right)
			\ln\Bigl[
						\left(
							1-\theta
						\right)
						^\frac{1}{1+\rho}
						+\theta^\frac{1}{1+\rho}
					\Bigr]
-\rho R
\Bigr\}
\label{ML error_exponent - 2.5}
\nonumber\\
& = & 
\max_{0\leq\rho\leq1}
\Bigl\{
			\rho \ln2
			-\left(1+\rho\right)
			\calV\left(\theta,\frac{1}{1+\rho}\right)
			-\rho R
\Bigr\}
\label{ML error_exponent - 2.75}
\nonumber\\
& \stackrel{\Delta}{=} &
\max_{0\leq\rho\leq1}
	E_{r}(\theta,\rho),
\end{eqnarray}
where in ($a$), the inner maximization is achieved by taking $Q^{*}=\{\frac{1}{2},\frac{1}{2}\}$ (\cite[Sect. 3.4]{VO79}).\\
Let us now define $\rho'=\frac{\lambda \rho}{1-\lambda \rho}$ and $\rho''=\frac{1}{\lambda}-1$, and rewrite the numerator of (\ref{BSC - xi for Q-star - 4}) as follows:

\begin{lefteqn}
{				\rho \ln2
				-\calV ( \theta,1- \lambda \rho )
				-\rho \calV \left(\theta',\lambda \right)
				-\rho R
=
}
\end{lefteqn}
\begin{eqnarray}
&=&
				\rho\ln 2
				-\calV \left(\theta,\frac{1}{1+\rho'}\right)
				-\rho\calV \left(\theta',\frac{1}{1+\rho''}\right)
				-\rho R
\nonumber\\
&=&
(1-\lambda \rho)
\Bigl[
	\rho'\ln 2
	-(1+\rho') \calV \left(\theta,\frac{1}{1+\rho'}\right)
	-\rho'R
\Bigr]
+
\nonumber\\
& &
\lambda \rho
\Bigl[
	\rho''\ln 2
	-(1+\rho'') \calV \left(\theta',\frac{1}{1+\rho''}\right)
	-\rho''R
\Bigr]
\nonumber\\
&=&
(1-\lambda \rho) \cdot E_{r} \left(\theta,\rho'\right)+
\lambda \rho \cdot E_{r} \left(\theta',\rho''\right)
\nonumber\\
&=&
(1-\lambda \rho) \cdot E_{r} \left(\theta,\frac{\lambda \rho}{1-\lambda \rho}\right)+
\lambda \rho \cdot E_{r} \left(\theta',\frac{1}{\lambda}-1\right).
\end{eqnarray}
Finally, we get that
\begin{eqnarray}
\label{BSC - xi for Q-star - 5}
\xi_{LB}^{*}\left(R\right)
& \geq &
\min_{\theta,\theta' \in \Theta}
\max_{\stackrel{0\leq\rho\leq1}{0\leq\lambda\leq 1/\rho}}
		\frac{
				(1-\lambda \rho) \cdot E_{r}(\theta,\frac{\lambda \rho}{1-\lambda \rho})+
				\lambda \rho \cdot E_{r}(\theta',\frac{1}{\lambda}-1)
		}
		{
			(1- \lambda \rho) \cdot E_{r}^{*}(\theta)
			+\lambda \rho \cdot E_{r}^{*}(\theta')
		}.
\end{eqnarray}
Now, by choosing $\lambda=\frac{1}{1+\tilde{\rho}}$, where $\tilde{\rho}$ is the achiever
of $E_{r}^{*}(\theta')=\max_{0\le\rho\le 1} E_{r}(\theta',\rho)$, 
and 
$\rho=
\frac{\hat{\rho}}{1+\hat{\rho}}
(1+\tilde{\rho})
$, where $\hat{\rho}$
is the achiever of $E_{r}^{*}(\theta)=\max_{0\le\rho\le 1}E_{r}(\theta,\rho)$ 
(observing that $
\frac{\hat{\rho}}{1+\hat{\rho}}
(1+\tilde{\rho})
\le 1$,
therefore this choice is feasible), we get that both the numerator and the denominator of (\ref{BSC - xi for Q-star - 5}) equal to 
$
				(1-\lambda \rho) \cdot E_{r}(\theta,\hat{\rho})+
				\lambda \rho \cdot E_{r}(\theta',\tilde{\rho})
$, and so, $\xi_{LB}^{*}\left(R\right)=1$.

We should note that for the BSC model, the same conclusion (i.e., $\xi^{*}=1$) holds also for linear codes and systematic linear codes (as the optimal random coding distribution that was used is $Q^{*}=\{\frac{1}{2},\frac{1}{2}\}$ (see (\ref{ML error_exponent - 2.5})).

\section{Proof of Theorem 3}
First, consider a given channel output related to the entire transmitted sequence of information. Without loss of generality, the all-zero message will be assumed to be transmitted. Let us now consider a segment of length $K+l$, $l \geq 0$, of the transmitted information vector, and any other incorrect path diverging from it at node $j$ and emerging at node $j+K+l$ (note that the minimum length of a diverging path is $K$ since after a non-zero vector is inserted to the encoder, $K-1$ zero vectors are needed in order to return to the all-zero state).

We observe that the information sequence related to such an incorrect path has the following structure (we ignore the values of the information sequence outside the range $(j, j+K+l-1)$):
$$
\buj, \bujplusone,\ldots, \bujplusl, \underbrace{\bzero,\ldots, \bzero}_{K-1}
$$ 
where all of the vectors are of length $b$.\\
In order for the incorrect path to diverge exactly from node $j$ to node $j+K+l$, $\buj$ and $\bujplusl$ can be any of the $2^b-1$ non-zero vectors (thus, there are $(2^b-1)^2$ possibilities for their values), and each of the $l-1$ information vectors $\bujplusone,\ldots, \bujpluslminusone$ can be any binary vector of length $b$, with the restriction of no more than $K-2$ consecutive all-zero vectors (thus, there are less than $2^{b\left(l-1\right)}$ possibilities for their values). Therefore, the number of such incorrect paths, denoted by $M$, is upper-bounded by
\begin{eqnarray}
\label{number_of_incorrect_paths}
M 
\leq
\left(2^b-1\right)^2 2^{b\left(l-1\right)}
\leq
\left(2^b-1\right) 2^{bl}
\end{eqnarray}

We next upper bound the probability that an incorrect path is preferred by the minimax decoder (minimizing the metric $\rho$) over the correct path, and then average this probability over the ensemble of time--varying convolutional codes.\\
We will use $\bVj=[\bvj,\bvjplusone,\ldots,\bvjplusKpluslminusone]$ to denote the code vector of length $N=n(K+l)$ that corresponds to the correct all-zeros path, while $\bVjtag$ and $\bVjtagaim$ will be used to denote code vectors that correspond to other incorrect paths. The notation $\bolVj$ will be used for the complement vector of $\bVj$. A segment of length $N$ of the corresponding channel output will be denoted by $\bYj$, and $Q^{*}$ will be used to denote the random coding distribution.

\begin{lefteqn}
{\overline{
	\Pr\left\{
		\rho(\bVjtag,\bYj)	
		\leq
		\rho(\bVj,\bYj)
			| \theta
		\right\}
	}
=
\nonumber
}
\end{lefteqn}
\begin{eqnarray}& = &
\sum_{\bVj,\bVjtag}
Q^{*}(\bVj,\bVjtag)
\Pr\left\{
		\rho(\bVjtag,\bYj)	
		\leq
		\rho(\bVj,\bYj)
			| \theta
		\right\}
\nonumber\\
& \stackrel{(a)}{=} &
2^{-2N}
\sum_{\bVj,\bVjtag}
\Pr\left\{
		\rho(\bVjtag,\bYj)	
		\leq
		\rho(\bVj,\bYj)
		\right\}
\nonumber\\
& = &
2^{-2N}
\sum_{\bVj,\bVjtag}
\Pr\Bigl\{
			\min\left\{
						\delta{(\bVjtag,\bYj)} ,
						 1-\delta{(\bVjtag,\bYj)}
					\right\}
			\leq
			\min\left\{
						\delta{(\bVj,\bYj)} , 1-\delta{(\bVj,\bYj)}
					\right\}								
		\Bigr\}
\nonumber\\
& = &
2^{-2N}
\sum_{\bVj,\bVjtag}
\Pr\Bigl\{
    \left[
			\delta{(\bVjtag,\bYj)}
			\leq
			\min\left\{
						\delta{(\bVj,\bYj)} , 1-\delta{(\bVj,\bYj)}
					\right\}								
		\right]
\nonumber\\
& &
\ \ \ \ \ \ \ \ \ \ \ \ \ \ \ \ \ \ 
\bigcup
		\left[
			\delta{(\bolVjtag,\bYj)}
			\leq
			\min\left\{
						\delta{(\bVj,\bYj)} , 1-\delta{(\bVj,\bYj)}
					\right\}								
		\right]
		\Bigr\}
\nonumber\\
& \stackrel{(b)}{\leq} &
2^{-2N}
\sum_{\bVj,\bVjtag}
\Pr\Bigl\{
			\delta{(\bVjtag,\bYj)}
			\leq
			\delta{(\bVj,\bYj)}
\bigcup
			\delta{(\bolVjtag,\bYj)}
			\leq
			\delta{(\bVj,\bYj)}
		\Bigr\}
\nonumber\\
& \stackrel{(c)}{\leq} &
2^{-2N}
\sum_{\bVj}
\sum_{\bVjtag}
\Pr\Bigl\{
			\delta{(\bVjtag,\bYj)}
			\leq
			\delta{(\bVj,\bYj)}
		\Bigr\}
+
2^{-2N}
\sum_{\bVj}
\sum_{\bVjtag}
\Pr\Bigl\{
			\delta{(\bolVjtag,\bYj)}
			\leq
			\delta{(\bVj,\bYj)}
	\Bigr\}
\nonumber\\
& \stackrel{(d)}{=} &
2^{-2N}
\sum_{\bVj}
\sum_{\bVjtag}
\Pr\Bigl\{
			\delta{(\bVjtag,\bYj)}
			\leq
			\delta{(\bVj,\bYj)}
		\Bigr\}
+
2^{-2N}
\sum_{\bVj}
\sum_{\bVjtagaim}
\Pr\Bigl\{
			\delta{(\bVjtagaim,\bYj)}
			\leq
			\delta{(\bVj,\bYj)}
	\Bigr\}
\nonumber\\
& = &
2\cdot2^{-2N}
\sum_{\bVj}
\sum_{\bVjtag}
\Pr\Bigl\{
			\delta{(\bVjtag,\bYj)}
			\leq
			\delta{(\bVj,\bYj)}
		\Bigr\}
\nonumber\\
& \stackrel{(e)}{\leq} &
2\cdot2^{-2N}
\sum_{\bVj}
\sum_{\bVjtag}
\sum_{\bYj}
\sqrt{P_{\theta}(\bYj|\bVj)P_{\theta}(\bYj|\bVjtag)}
\nonumber\\
& \stackrel{(f)}{=} &
2\left\{
	\sum_{y}
		\left[
			\sum_{v}
			\frac{1}{2}\sqrt{P_{\theta}(y|v)}
		\right]^2
\right\}^N
\nonumber\\
\label{segment error upper bound}
& {\stackrel{\cdot}{=}} &
e^{-N 
   R_{\theta,0}
   (Q^{*})
   },
\end{eqnarray}
where
$$
 R_{\theta,0}
   \Bigl(
   		Q^{*}
    \Bigr)
 =    
 -\ln
 \sum_{y}
		\left[
			\sum_{v}
			\frac{1}{2}\sqrt{P_{\theta}(y|v)}
		\right]^2.
$$
In ($a$) we used the fact that both $\bVj$ and $\bVjtag$ can attain each of their $2^N$ possible values equiprobably and independently.
This claim for $\bVj$ (which corresponds to the all--zero path) can be justified due to the fact that the elements of $\bGjt, 0 \leq j \leq K-1$ and $\bvzerot$ are repeatedly randomized at each time instant (see (\ref{ensemble of time-varying convolutional codes definition})). Therefore, $\forall \  0\leq i \leq K+l, \bvjplusi=\bvzerojplusi$, thus each one of these vectors is likely to attain each of its $2^n$ values equiprobably.
This claim for $\bVjtag$ (which correspond to the incorrect path) can be justified since $\buj$ and $\bujplusl$ are non--zero and $\bujplusone,\ldots, \bujpluslminusone$ cannot include more than $K-2$ consecutive all--zero vectors. Thus, each code vector of $\bvtagjplusi$, $0\leq i \leq K+l$ is formed by the modulo-2 sum of $\bvzerojplusi$ with at least one of the rows of $\bGzerojplusi, \bGonejplusi,\ldots,\bGKminusonejplusi$ and is therefore likely to attain each of its $2^n$ values with equal probability as well and independently with the other code vectors (this fact is dealt in details in \cite[Sect. 5.1]{VO79}).
($b$) is true since we switched into looser conditions inside each event in the probability term. In ($c$) we used the union bound.
In ($d$) we used the fact that observing $\delta{(\bolVj,\bYj)}$, when  summing up over all of $\bVj$'s possible values, is equivalent to observing $\delta{(\bVj,\bYj)}$ (since in both cases, each of the $2^N$ values of the vector is covered by the summation).
In ($e$) we used the Bhattacharyya bound for the pairwise error probability when using ML decision rule, and ($f$) is true since the channel is memoryless.

We proved that the probability that other code segment would be preferred by the minimax decoder over the correct segment, when averaged over the ensemble of time-varying convolutional codes, is upper bounded by twice the bound achieved for ML decoder in \cite{VO79}. Thus, it is exponentially of the same order. The subsequent steps in deriving an upper bound to the bit error exponent for rates
$R \leq R_{\theta,0}
			\left(Q^{*}\right)$
are identical to that of ML decoder (see \cite[Sect. 5.1]{VO79}) and the final result is the same.

Therefore, it was proved that when using the minimax decoder, the achievable exponent for bit error probability is no less than when the channel parameter is known and the ML decoder is used. The same error exponent was proved to be achievable for rates up to 
$
R_{\theta,0}
			\left(Q^{*}\right)
$. 

In order to extend the average upper bound for the bit error probability to rates higher than 
$
R_{\theta,0}
			\left(Q^{*}\right)
$, 
we will use a slightly different technique.\\
First, we upper bound $\pi_{l,\theta}(j)$, the probability that a branch in the minimax based decoding path will occur by any one of the other possible paths, starting at node $j$ and reemerging after $K+l$ branches. We should observe, as mentioned in (\ref{number_of_incorrect_paths}), that the number of such diverging paths satisfies $M \leq \left(2^b-1\right) 2^{bl}$. The code segments associated with these $M$ incorrect paths will be denoted by $\bVjupone,\ldots,\bVjupM$, respectively.

\begin{lefteqn}
{
\pi_{l,\theta}(j) =
\nonumber
}
\end{lefteqn}
\begin{eqnarray}
& = &
\Pr\left\{
		\exists \  1 \leq i \leq M: 
		\rho(\bVjupi,\bYj)	
		\leq
		\rho(\bVj,\bYj)
	\right\}
\nonumber\\
& = &
\Pr\Bigl\{
\exists \   1 \leq i \leq M:
			\min\left\{
						\delta(\bVjupi,\bYj) , 1-\delta(\bVjupi,\bYj)
					\right\}
			\leq
			\min\left\{
						\delta(\bVj,\bYj) , 1-\delta(\bVj,\bYj)
					\right\}
	\Bigr\}
\nonumber\\
& = &
\Pr\Bigl\{
\exists \   1 \leq i \leq M: 
			\delta(\bVjupi,\bYj)
			\leq
			\min\left\{
						\delta(\bVj,\bYj) , 1-\delta(\bVj,\bYj)
					\right\}								
			\bigcup
		\nonumber\\
& &
\ \ \ \ \ \ \ \ \ \ \ \ \ \ \ \ \ \ \ \ \ \ \
			\delta(\bolVji,\bYj)
			\leq
  		\min\left\{
						\delta(\bVj,\bYj) , 1-\delta(\bVj,\bYj)
					\right\}								
	\Bigr\}
\nonumber\\
& \stackrel{(a)}{\leq} &
\Pr\Bigl\{
\exists \   1 \leq i \leq M: 
			\delta(\bVjupi,\bYj)
			\leq
			\delta(\bVj,\bYj)
			\bigcup
			\delta(\bolVji,\bYj)
			\leq
						\delta(\bVj,\bYj)
	\Bigr\}
\nonumber\\
& \stackrel{(b)}{\leq} &
\Pr\Bigl\{
\exists \   1 \leq i \leq M: 
			\delta(\bVjupi,\bYj)
			\leq
			\delta(\bVj,\bYj)
\bigr\}
+
\nonumber\\
& &
\Pr\Bigl\{
\exists \   1 \leq i \leq M: 
			\delta(\bolVji,\bYj)
			\leq
						\delta(\bVj,\bYj)
	\Bigr\}
\nonumber\\
& \stackrel{(c)}{\leq} &		
\label{pie l theta upper bound}
\sum_{\bYj}
	P_{\theta}\bigl(\bYj|\bVj\bigr)
	^{\frac{1}{1+\rho}}
\Bigl[
\sum_{i=1}^{M}
	P_{\theta}\bigl(\bYj|\bVjupi\bigr)
	^{\frac{1}{1+\rho}}	
\Bigr]^\rho
+
\sum_{\bYj}
	P_{\theta}\bigl(\bYj|\bVj\bigr)
	^{\frac{1}{1+\rho}}
\Bigl[
\sum_{i=1}^{M}
	P_{\theta}\bigl(\bYj|\bolVji\bigr)
	^{\frac{1}{1+\rho}}	
\Bigr]^\rho,
\end{eqnarray}
where (a) is true since we increased the right terms of the two inequalities, and thus increased the probability for union of these two events,
in ($b$),  the union bound was used,
and in ($c$), we used the Gallager bound for the error probability when using the ML decision rule. This error was used for each of the two error probabilities.

We now move to upper bound the average of $\pi_{l,\theta}(j)$ over the ensemble of time-varying colvolutional codes:

\begin{lefteqn}
{
\overline{\pi_{l,\theta}(j)} = 
}
\end{lefteqn}
\begin{eqnarray}
& \stackrel{(a)}{=} &
\sum_{\bVj}
2^{-N}
\sum_{\bVjupone}
\ldots
\sum_{\bVjupM}
2^{-NM}
\pi_{l,\theta}(j)
\nonumber\\
& \stackrel{(b)}{\leq} &		
\sum_{\bYj}
\sum_{\bVj}
2^{-N}
	P_{\theta}\bigl(\bYj|\bVj\bigr)
	^{\frac{1}{1+\rho}}
\sum_{\bVjupone}
\ldots
\sum_{\bVjupM}
2^{-NM}
\nonumber\\
& &
\biggl[
\Bigl[
\sum_{i=1}^{M}
	P_{\theta}\bigl(
								\bYj|\bVjupi
						\bigr)
	^{\frac{1}{1+\rho}}
\Bigr]^\rho
+
\Bigl[
\sum_{i=1}^{M}
	P_{\theta}\bigl(\bYj|\bolVji\bigr)
	^{\frac{1}{1+\rho}}	
\Bigr]^\rho
\biggr]
\nonumber\\
& \stackrel{(c)}{=} &		
2\cdot
\sum_{\bYj}
\sum_{\bVj}
2^{-N}
	P_{\theta}\bigl(\bYj|\bVj\bigr)
	^{\frac{1}{1+\rho}}
\sum_{\bVjupone}
\ldots
\sum_{\bVjupM}
2^{-NM}
\Bigl[
\sum_{i=1}^{M}
	P_{\theta}\bigl(
								\bYj|\bVjupi
						\bigr)
	^{\frac{1}{1+\rho}}
\Bigr]^\rho
\nonumber\\
& \stackrel{(d)}{\leq} &		
2\cdot
\sum_{\bYj}
\sum_{\bVj}
2^{-N}
	P_{\theta}\bigl(\bYj|\bVj\bigr)
	^{\frac{1}{1+\rho}}
\Bigl[	
	\sum_{i=1}^{M}
	\sum_{\bVjupone}
	\ldots
	\sum_{\bVjupM}
	2^{-NM}
	P_{\theta}\bigl(
								\bYj|\bVjupi
						\bigr)
	^{\frac{1}{1+\rho}}
\Bigr]^\rho
,
\ 0 \leq \rho \leq 1
\nonumber\\
& \stackrel{(e)}{=} &		
2\cdot
\sum_{\bYj}
\sum_{\bVj}
2^{-N}
	P_{\theta}\bigl(\bYj|\bVj\bigr)
	^{\frac{1}{1+\rho}}
\Bigl[	
	\sum_{i=1}^{M}
	\sum_{\bVjupi}
	2^{-N}
	P_{\theta}\bigl(
								\bYj|\bVjupi
						\bigr)
	^{\frac{1}{1+\rho}}
\Bigr]^\rho
,\ \ \ \ \ 0 \leq \rho \leq 1
\nonumber\\
& \stackrel{(f)}{\leq} &		
2\cdot
\left(2^b-1\right) 2^{bl\rho}
\sum_{\bYj}
\sum_{\bVj}
2^{-N}
	P_{\theta}\bigl(\bYj|\bVj\bigr)
	^{\frac{1}{1+\rho}}
\Bigl[	
	\sum_{\bVjupi}
	2^{-N}
	P_{\theta}\bigl(
								\bYj|\bVjupi
						\bigr)
	^{\frac{1}{1+\rho}}
\Bigr]^\rho
,\ \ \ \ \ 0 \leq \rho \leq 1
\nonumber\\
& = &		
2\cdot
\left(2^b-1\right) 2^{bl\rho}
\sum_{\bYj}
\Bigl[	
	\sum_{\bVj}
	2^{-N}
	P_{\theta}\bigl(
								\bYj|\bVj
						\bigr)
	^{\frac{1}{1+\rho}}	
\Bigr]^{1+\rho}
,
\ \ \ \ 0 \leq \rho \leq 1
\nonumber\\
& \stackrel{(g)}{=} &		
2\cdot
\left(2^b-1\right) 2^{bl\rho}
\left\{
\sum_{y}
\Bigl[	
	\sum_{v}
	2^{-N}
	P_{\theta}\left(
								y|v
						\right)
	^{\frac{1}{1+\rho}}
\Bigr]^{1+\rho}
\right\}^N
,
\ \ \ \ \ 0 \leq \rho \leq 1
\nonumber\\
\label{segment error upper bound - high rate}
& \stackrel{\cdot}{=} &
\left(2^b-1\right) 2^{bl\rho}
e^{-\left(K+l\right)n
   E_{\theta,0}
   (\rho,
   	\left\{\frac{1}{2},
           \frac{1}{2}
    \right\})
   }
,
\ \ \ \ 
0 \leq \rho \leq 1,
\end{eqnarray}
where
$$
 E_{\theta,0}
   \left(
   	\rho,
   	Q^{*}
   \right)
 =    
 -\ln
\sum_{y}
	\Bigl[	
		\sum_{v}
			2^{-N}
			P_{\theta}\left(
									y|v
								\right)
			^{\frac{1}{1+\rho}}
	\Bigr]^{1+\rho}.
$$
In ($a$), we sum over all possible code vectors associated with the different paths in the trellis. As explained earlier, each code vector can attain all of its $2^N$ values equiprobably and independently with the other code vectors.
In ($b$), we used the result from (\ref{pie l theta upper bound}).
($c$) is true since examining 
$P_{\theta}\bigl(\bYj|\bolVji\bigr), 1\leq i \leq M$, when summing up over all of $\bVjupone,\ldots,\bVjupM$ possible values, is equivalent to the examination of 
$P_{\theta}\bigl(\bYj|\bVjupi\bigr), 1\leq i \leq M$
.
In ($d$), we bound ourselves to $0 \leq \rho \leq 1$ and use Jensen's inequality.
($e$) is true since for a fixed $i$, $P_{\theta}\bigl(\bYj|\bVjupi\bigr)$ depends only on $\bVjupi$, and is enumerated for the $2^{N(M-1)}$ possibilities of 
$
\bVjupone,\ldots,
\bVjupiminusone,
\bVjupiplusone,\ldots,
\bVjupM
$.
In ($f$), we upper bound $M$ by $\left(2^b-1\right) 2^{bl}$, and ($g$) is true since the BSC is memoryless.

As in the above proof for rates up to 
$R_{\theta,0}
			\left(Q^{*}\right)$
, the subsequent steps in deriving an upper bound to the bit error exponent for rates
$R_{\theta,0}
			\left(Q^{*}\right)
\leq
R
\leq
C_{\theta}
$
for the minimax decoder are identical to that of ML decoder (see \cite[Sect. 5.1]{VO79}) and the final result is the same.
This completes the proof that the achievable exponent for bit error probability of the minimax decoder is equal to that of the ML decoder, for all rates up to capacity.

\section*{A. Appendix}
\subsection*{A.1 Proof of eq. (\ref{xi star - lower bound}) for ensembles of Linear and Systematic Linear Codes}
In this section, we examine the performance of the minimax decoding rule with respect to uniform i.i.d. random coding over ensembles of linear codes and systematic linear codes. We will prove that for a family of BIOS channels, the same single--letter formula for the lower bound to the achievable fraction $\xi^{*}$ is obtained, with uniform i.i.d. random coding distribution $Q^{*}=\left\{\frac{1}{2},\frac{1}{2}\right\}$ (i.e. $\Delta^*(P)=\ln 2-H(P)$).

Using Gallager's techniques, we first upper bound the decoding error probability given that the \textit{m}-th message was sent for a given \(\theta\) in the following way:
\begin{eqnarray}
P_{E_m}\left(
						\Omega|\theta
			 \right)
&=&
\sum_{\by\in {\cal Y}^N}
			P_\theta(
									\by|\bvm
							)
			\textsl{1}\Bigl\{
			\exists m'\neq m :
					\max_{\theta' \in \Theta_{N}}
					f_{\theta'}(
												\bvmtag,\by
											)
					\geq
					\max_{\theta'' \in \Theta_{N}}f_{\theta''}(
																				\bvm,\by
																		)
			 \Bigr\}
\nonumber\\
&=&
\sum_{\by\in {\cal Y}^N}
			P_\theta(
								\by|\bvm
							)
			\textsl{1}\Bigl\{
						\exists \theta',\exists m'\neq m :
						f_{\theta'}(
													\bvmtag,\by
											 )
						\geq
						\max_{\theta'' \in \Theta_{N}}f_{\theta''}(
																						\bvm,\by
																				)
			\Bigr\}
\nonumber\\
& \stackrel{(a)}{=}&
	\sum_{\by\in {\cal Y}^N}
			P_\theta(
									\by|\bvm
							)
			\max_{\theta' \in \Theta_{N}}\textsl{1}\Bigl\{
												\exists m'\neq m:
												f_{\theta'}(
																			\bvmtag,\by
																	)
												\geq
												\max_{\theta'' \in \Theta_{N}}f_{\theta''}
												(
														\bvm,\by
												)
										\Bigr\}
\nonumber\\
&=&\sum_{\by\in {\cal Y}^N}
	P_\theta(
							\by|\bvm
					)
	\max_{\theta' \in \Theta_{N}}\textsl{1}\Bigl\{
																	\exists m'\neq m:
																	\frac{f_{\theta'}(
																											\bvmtag,\by
																										)}
																	{\max_{\theta'' \in \Theta_{N}}
																	f_{\theta''}(
																									\bvm,\by
																								)}\geq1\Bigr\}
\nonumber\\
&\stackrel{(b)}{\leq}&	
\sum_{\by\in {\cal Y}^N}
	P_\theta(
						\by|\bvm
				)
\max_{\theta' \in \Theta_{N}}
\min_{\stackrel{\lambda\geq0}{\rho\geq0}}
\Biggl[
					\sum\limits_{m'\neq m}\Bigl(
 																	\frac{e^{Nf_{\theta'}(
																													\bvmtag,\by
 																												)}}
																	{\max_{\theta'' \in \Theta_{N}}
																		e^{Nf_{\theta''}(
																												\bvm,\by
																										 )}}
																\Bigr)^\lambda
\Biggr]^\rho,
\label{m-message error prob. upper bound}
\end{eqnarray}
where $\rho \geq 0$ and $\lambda \geq 0$ are free parameters. \\
($a$) is true since if we denote with $A(\theta)$ an event dependent on $\theta \in \Theta_{N}$, and denote with $C$ a constant, then 
$$
\textsl{1}\left\{
						\exists \theta \in \Theta_{N}: A(\theta) > C
			\right\}
=
\max_{\theta \in \Theta_{N}}\textsl{1}\left\{A(\theta) > C\right\}.
$$
($b$) is true since if we denote with $f_{1}(m')$ and $f_{2}(m)$ two non--negative functions of $m'$ and $m$ respectively, then (using Gallager's technique)
$$
\textsl{1}\left\{
						\exists m' \neq m: \frac{f_{1}(m')}{f_{2}(m)} \geq 1
			\right\}
\leq
\min_{\stackrel{\lambda\geq0}{\rho\geq0}}
\left[
\sum_{m' \neq m} \left(
											\frac{e^{f_{1}(m')}}{e^{f_{2}(m)}}
									\right)^\lambda
\right]^\rho.							
$$
Based on (\ref{m-message error prob. upper bound}), we now develop an upper bound to the minimax criterion related to a specific linear code (i.e., specific values of $\bG$ and $\bvzero$, thus denoted by $S_N\left(\bvzero,\bG\right)$):
\begin{eqnarray}
\label{LC S_N upper bound_1}
S_N\left(\bvzero,\bG\right)&=&
\max_{\theta \in \Theta}
			\Bigl\{
				\frac
					{P_{E}\left(\Omega|\theta\right)}
					{[{\overline{P}_{E}}^{*}\left(\theta\right)]^\xi}
			\Bigr\}
\stackrel{\cdot}{=}
\max_{\theta \in \Theta}
			\Bigl\{
				\frac
					{P_{E}\left(\Omega|\theta\right)}
					{e^{-N\xi E_{r}^{*}(\theta)}}
			\Bigr\}		
\nonumber\\			
\label{LC S_N upper bound_2}
&\leq &
\max_{\theta \in \Theta}
			\Biggl\{
				\frac{1}{M}\sum\limits_{m=0}^{M-1}
								\sum_{\by\in {\cal Y}^N}
								{e^{N\xi E_{r}^{*}(\theta)}}
								P_\theta(\by|\bvm)
								\max_{\theta' \in \Theta_N}
								\min_{\stackrel{\lambda\geq0}{\rho\geq0}}
\nonumber\\								
& &	
\ \ \ \ \ \ 
								\left(
									\sum\limits_{m'\neq m}\left[
																					\frac{e^{Nf_{\theta'}\left(
																																\bvmtag,\by
 																															\right)}}
																					{\max_{\theta'' \in \Theta_N}
																					e^{Nf_{\theta''}\left(
																															\bvm,\by
																													 \right)}}
																				\right]^\lambda\right)^\rho
			\Biggr\}		
\nonumber\\
\label{LC S_N upper bound_3}
&=&
\max_{\theta \in \Theta}
			\Biggl\{
				\frac{1}{M}\sum\limits_{m=0}^{M-1}
								\sum_{\by\in {\cal Y}^N}
								{e^{N\cdot f_{\theta}(\bvm,\by)}}
								\max_{\theta' \in \Theta_N}
								\min_{\stackrel{\lambda\geq0}{\rho\geq0}}
\nonumber\\								
& &	
\ \ \ \ \ \ 
								\left(
									\sum\limits_{m'\neq m}\left[
																					\frac{e^{Nf_{\theta'}\left(
																																\bvmtag,\by
 																															\right)}}
																					{\max_{\theta'' \in \Theta_N}
																					e^{Nf_{\theta''}\left(
																															\bvm,\by
																													 \right)}}
																				\right]^\lambda\right)^\rho
			\Biggr\}		
\nonumber\\		
\label{LC S_N upper bound_4_1}
& \stackrel{(a)}{\leq} &
	\frac{1}{M}\sum\limits_{m=0}^{M-1}
				\sum_{\by\in {\cal Y}^N}
						\Bigl(
							\max_{\theta \in \Theta_N}
								e^{N {f_{\theta}(\bvm,\by)}}
						\Bigr)
						\max_{\theta' \in \Theta_N}
						\min_{\stackrel{\lambda\geq0}{\rho\geq0}}
								\left\{
									\frac
										{
										\biggl[
											\sum\limits_{m'\neq m}
												{e^{N \lambda f_{\theta'}
													\left(
														\bvmtag,\by
													\right)}}
										\biggr]^\rho
										}
										{\left(
													\max_{\theta'' \in \Theta_N}
														{e^{N{f_{\theta''}(\bvm,\by)}}}
										\right)^{\lambda\rho}}
								\right\}
\nonumber\\
\label{LC S_N upper bound_5}
& \stackrel{(b)}{\leq} &
	\frac{1}{M}\sum\limits_{m=0}^{M-1}
			\sum_{\by\in {\cal Y}^N}
				\max_{\theta' \in \Theta_N}
				\min_{\stackrel{\lambda\geq0}{\rho\geq0}}
						\Biggl\{					
								\Bigl[
									\max_{\theta \in \Theta_N}
										{e^{N{f_{\theta}(\bvm,\by)}}}
								\Bigr]^{1-\lambda\rho}
								\biggl[
											\sum\limits_{m'\neq m}
												{e^{N \lambda f_{\theta'}
												\left(
														\bvmtag,\by
												\right)}}
								\biggr]^\rho\
						\Biggr\}
\nonumber\\
\label{LC S_N upper bound_6}
& \stackrel{(c)}{\leq} &
	\frac{1}{M}\sum\limits_{m=0}^{M-1}
			\sum_{\by\in {\cal Y}^N}
				\max_{\theta' \in \Theta_N}
				\min_{\stackrel{\rho\geq0}{0\leq\lambda\leq 1/\rho}}
						\max_{\theta \in \Theta_N}
							\Biggl\{					
										{e^{N({1-\lambda\rho})
											{f_{\theta}(\bvm,\by)}}}
								\biggl[
											\sum\limits_{m'\neq m}
												{e^{N \lambda f_{\theta'}
												\left(
														\bvmtag,\by
												\right)}}
								\biggr]^\rho\
							\Biggr\}
\nonumber\\
\label{LC S_N upper bound_7}
& \stackrel{(d)}{=} &
	\frac{1}{M}\sum\limits_{m=0}^{M-1}
			\sum_{\by\in {\cal Y}^N}
					\max_{\theta\in \Theta_N}
						\max_{\theta'\in \Theta_N}\min_
							{\stackrel{\rho\geq0}{0\leq\lambda\leq 1/\rho}}
							\Biggl\{					
										{e^{N({1-\lambda\rho})
											{f_{\theta}(\bvm,\by)}}}
								\biggl[
											\sum\limits_{m'\neq m}
												{e^{N \lambda f_{\theta'}
												\left(
														\bvmtag,\by
												\right)}}
								\biggr]^\rho\
							\Biggr\}.
\nonumber\\
\end{eqnarray}
The passages ($a$)--($d$) are explained as follows: In ($a$) we used the fact that the maximum of an expectation is no greater than the expectation of the maximum and changed the maximization of $\theta$ to be over $\Theta_N$. ($b$) is true since $\theta$ and ${\theta}''$ maximize two identical expressions, and therefore can be united. In ($c$) we restricted the range of the optimization to $1-\lambda\rho\geq0$ $(\Rightarrow\lambda\leq 1/\rho)$. In ($d$) we used the fact that for given $\bvm$, $\by$ and $\theta'$
\begin{eqnarray}
				\min_{\stackrel{\rho\geq0}{0\leq\lambda\leq 1/\rho}}
				\max_{\theta \in \Theta_N}
							\Biggl\{					
										{e^{N({1-\lambda\rho})
											{f_{\theta}(\bvm,\by)}}}
								\biggl[
											\sum\limits_{m'\neq m}
												{e^{N \lambda f_{\theta'}
												\left(
														\bvmtag,\by
												\right)}}
								\biggr]^\rho\
							\Biggr\}
& = &
\nonumber\\
						\max_{\theta\in \Theta_N}
						\min_{\stackrel{\rho\geq0}{0\leq\lambda\leq 1/\rho}}
							\Biggl\{					
										{e^{N({1-\lambda\rho})
											{f_{\theta}(\bvm,\by)}}}
								\biggl[
											\sum\limits_{m'\neq m}
												{e^{N \lambda f_{\theta'}
												\left(
														\bvmtag,\by
												\right)}}
								\biggr]^\rho\
							\Biggr\}.
\label{LC S_N inner optimization equivalence}
\end{eqnarray}
This interchange between the minimization over $\lambda$ and $\rho$ and the maximization over $\theta$ is justified in the Appendix, Section A.2.

Prior to deriving the single--letter formula for the lower bound to $\xi^{*}$, we first present the following claim:
\begin{lemma}
When a linear code is used for a BIOS channel and minimax decoding is used, the error probability for the $m$-th message is equal for all $0 \leq m \leq M-1$.
\end{lemma}
This lemma is proved in Section A.4 of the Appendix.

Based on this observation, we can assume, without loss of generality, that $\buzero=\bzero$ was transmitted, and then the upper bound to $S_N$ can be expressed as:
\begin{eqnarray}
\label{LC S_N upper bound-updated}
S_N\left(\bvzero,\bG\right)
& \leq &
			\sum_{\by \in Y^N}
					\max_{\theta \in \Theta_N}
						\max_{\theta' \in \Theta_N}\min_
							{\stackrel{\rho\geq0}{0\leq\lambda\leq 1/\rho}}
								\biggl\{					
									{e^{N ({1-\lambda\rho}) {f_{\theta}(\bvzero,\by)}	}}
									\Bigl[
											\sum\limits_{m=1}^{M-1}
													{e^{N \lambda f_{\theta'}
													\left(
																\bvm,\by
														\right)}}
										\Bigr]^\rho\
								\biggr\}
\nonumber\\					
\end{eqnarray}

In the following subsections, we will use the same technique to derive two upper bounds on the minimax criterion, one for the ensemble of linear codes and one for the ensemble of systematic linear codes.

\subsubsection*{Linear Codes}

By averaging $S_N$ over the ensemble of linear codes:
\begin{eqnarray}
\label{LC E[S_N] upper bound_1}
\overline{S}_N
&\stackrel{(a)}{=}&
		2^{-\left(K+1\right)N}
		\sum\limits_{\bvzero,\bG}
		S_N\left(\bvzero,\bG\right)
\nonumber\\
\label{LC E[S_N] upper bound_2_A}
&\leq&
2^{-\left(K+1\right)N}
\sum\limits_{\bvzero,\bG}
		\sum_{\by \in Y^N}
					\max_{\theta \in \Theta_N}
						\max_{\theta' \in \Theta_N}\min_
							{\stackrel{\rho\geq0}{0\leq\lambda\leq 1/\rho}}
								\Biggl\{					
									{e^{N ({1-\lambda\rho}) {f_{\theta}(\bvzero,\by)} }}
									\Biggl[
											\sum\limits_{m=1}^{M-1}
															{e^{N \lambda f_{\theta'}
													\left(
																\bvm,\by
														\right)}}
										\Biggr]^\rho\
								\Biggr\}
\nonumber\\								
\\
\label{LC E[S_N] upper bound_2_1}
&\stackrel{(b)}{\leq}&
2^{-\left(K+1\right)N}
\sum\limits_{\bvzero,\bG}
			\sum_{\by \in Y^N}
					\sum_{\theta \in \Theta_N}
						\sum_{\theta' \in \Theta_N}\min_
							{\stackrel{\rho\geq0}{0\leq\lambda\leq 1/\rho}}
								\Biggl\{					
									{e^{N ({1-\lambda\rho}) {f_{\theta}(\bvzero,\by)} }}
									\Biggl[
											\sum\limits_{m=1}^{M-1}
													{e^{N \lambda f_{\theta'}
													\left(
																\bvm,\by
														\right)}}
										\Biggr]^\rho\
								\Biggr\}
\nonumber\\
\label{LC E[S_N] upper bound_3_1}
&\stackrel{(c)}{\leq}&
			\sum_{\by \in Y^N}
					\sum_{\theta \in \Theta_N}
						\sum_{\theta' \in \Theta_N}
						\min_{\stackrel{\rho\geq0}{0\leq\lambda\leq 1/\rho}}
\nonumber\\					
& &											
								\Biggl\{					
									2^{-\left(K+1\right)N}
									\sum\limits_{\bvzero,\bG}
									\Biggl(
										{e^{N ({1-\lambda\rho}) {f_{\theta}(\bvzero,\by)} }}
										\Biggl[
											\sum\limits_{m=1}^{M-1}
													{e^{N \lambda f_{\theta'}
													\left(
																\bvm,\by
													\right)}}
										\Biggr]^\rho
									\Biggr)
								\Biggr\}
\nonumber\\
\label{LC E[S_N] upper bound_3_2_A}
&\stackrel{(d)}{\leq}&
\left| \Theta_N \right|^2
			\sum_{\by \in Y^N}
					\max_{\theta \in \Theta_N}
						\max_{\theta' \in \Theta_N}\min_
							{\stackrel{\rho\geq0}{0\leq\lambda\leq 1/\rho}}
\nonumber\\					
\label{LC E[S_N] upper bound_3_2_B}
& &																		
								\Biggl\{					
									2^{-N}
									\sum\limits_{\bvzero}
										{e^{N ({1-\lambda\rho}) {f_{\theta}(\bvzero,\by)} }}
											2^{-KN}
											\sum\limits_{\bG}										
												\Biggl[
													\sum\limits_{m=1}^{M-1}
															{e^{N \lambda f_{\theta'}
																\left(
																		\bvm,\by
																\right)}}
												\Biggr]^\rho
								\Biggr\}
\\
\label{LC E[S_N] upper bound_8}
&\stackrel{(e)}{\leq}&
\left| \Theta_N \right|^2
			\sum_{\by \in Y^N}
					\max_{\theta \in \Theta_N}
						\max_{\theta' \in \Theta_N}
						\min_{\stackrel{0\leq\rho\leq1}{0\leq\lambda\leq 1/\rho}}
\nonumber\\					
& &											
								\Biggl\{					
									2^{-N}
									\sum\limits_{\bvzero}
										{e^{N ({1-\lambda\rho}) {f_{\theta}(\bvzero,\by)} }}
										\Biggl[
											2^{-KN}
											\sum\limits_{\bG}										
													\sum\limits_{m=1}^{M-1}
															{e^{N \lambda f_{\theta'}
																\left(
																		\bvm,\by
																\right)}}
										\Biggr]^\rho
								\Biggr\}
\nonumber\\
\label{LC E[S_N] upper bound_9}
&=&
\left| \Theta_N \right|^2
			\sum_{\by \in Y^N}
					\max_{\theta \in \Theta_N}
						\max_{\theta' \in \Theta_N}
						\min_{\stackrel{0\leq\rho\leq1}{0\leq\lambda\leq 1/\rho}}
\nonumber\\
& &																	
								\Biggl\{					
									2^{-N}
									\sum\limits_{\bvzero}
										{e^{N ({1-\lambda\rho}) {f_{\theta}(\bvzero,\by)} }}
										\Biggl[
											2^{-KN}
												\sum\limits_{m=1}^{M-1}
													\sum\limits_{\bG}										
															{e^{N \lambda f_{\theta'}
																\left(
																		\bvm,\by
																\right)}}
										\Biggr]^\rho
								\Biggr\}
\nonumber\\					
\label{LC E[S_N] upper bound_10}
&\stackrel{(f)}{=}&
\left| \Theta_N \right|^2
			\sum_{\by \in Y^N}
					\max_{\theta \in \Theta_N}
						\max_{\theta' \in \Theta_N}
						\min_{\stackrel{0\leq\rho\leq1}{0\leq\lambda\leq 1/\rho}}
\nonumber\\
& &																	
								\Biggl\{					
									2^{-N}
									\sum\limits_{\bvzero}
										{e^{N ({1-\lambda\rho}) {f_{\theta}(\bvzero,\by)} }}
										\Biggl[
											\left(
												M-1
											\right)
											2^{-N}
												\sum\limits_{\bv}
													{e^{N \lambda f_{\theta'}
													\left(
															\bv
															,\by
													\right)}}
										\Biggr]^\rho
								\Biggr\}
\nonumber\\
\label{LC E[S_N] upper bound_12}
&\leq&
\left| \Theta_N \right|^2
			\sum_{\by \in Y^N}
					\max_{\theta \in \Theta_N}
						\max_{\theta' \in \Theta_N}
						\min_{\stackrel{0\leq\rho\leq1}{0\leq\lambda\leq 1/\rho}}
\nonumber\\
& &
								\Biggl\{
									M^\rho
									\Biggl[
										2^{-N}
										\sum\limits_{\bv}
											{e^{N ({1-\lambda\rho}) {f_{\theta}(\bv,\by)} }}
									\Biggr]
									\cdot
									\Biggl[
												2^{-N}
													\sum\limits_{\bvtag}
														{e^{N \lambda f_{\theta'}
														\left(
																\bvtag
																,\by
														\right)}}
										\Biggr]^\rho
									\Biggr\},
\nonumber\\
\label{LC E[S_N] upper bound_13}
\end{eqnarray}
where the steps ($a$)--($f$) are as follows: The equality in ($a$) is obtained by averaging over $2^{\left(K+1\right)N}$ equiprobable values of $\bvzero$ and $\bG$.
($b$) and ($d$) follow from the fact that for a non-negative function $f(\theta)$,
non-negative function $f(\theta)$, 
\begin{eqnarray}
	\max_{\theta \in \Theta_N} f(\theta)
	\leq
	\sum\limits_{\theta \in \Theta_N} f(\theta)
	\leq
	\left| \Theta_N \right| \cdot \max_{\theta \in \Theta_N} f(\theta).
\label{non negative f_theta inequalities}
\end{eqnarray}
($c$) is true since an expectation of a minimum is upper-bounded by the minimum of the expectation. In ($e$), we limit the optimization over $\rho$ to
$
0\leq{\rho}\leq1
$
and use Jensen's inequality. 
In ($f$), we used the following equivalence for the two inner summations:
\begin{eqnarray}
\label{LC E[S_N] inner term equivalence}
\sum\limits_{m=1}^{M-1}
\sum\limits_{\bG}										
	{e^{N \lambda f_{\theta'}
		\left(
				\bvm,\by
		\right)}}
=
(M-1)
2^{(K-1)N}
\sum\limits_{\bv}
	{e^{N \lambda f_{\theta'}
		\left(
				\bv
				,\by
		\right)}}.
\end{eqnarray}
This equivalence is proved in Section A.5 of the Appendix.

From (\ref{f dep on x n y-5}), we conclude that the term inside the summation in (\ref{LC E[S_N] upper bound_13}) is identical for all $\by$'s of the same type class. Thus, the summation can be conducted over types. 
Using (\ref{E upper bound}), we continue to upper bound $\overline{S}_N$ in the following way (note that the function $A(\theta,\alpha,P_{\bx\by})$ used here corresponds to a binary i.i.d. random coding distribution, as specified in (\ref{A_theta_for i.i.d.})):
\begin{eqnarray}
\label{E[S_N] upper bound_8_1}
\overline{S}_N
& \stackrel{(a)}{\leq} &
\left| \Theta_N \right|^2
			\sum\limits_{T_{\by}}
				\max_{\theta \in \Theta_N}
				\max_{\theta'\in \Theta_N}
					\min_{\stackrel{0\leq\rho\leq1}{0\leq\lambda\leq 1/\rho}}
\nonumber\\
& &										
	\Biggl\{					
			e^{N \rho R}
			e^{N H_{\by}(Y)}
			e^{N
				\left[
					(1- \lambda \rho) {\xi E_{r}^{*}(\theta)}
					-
					\min_{P_{\bx|\by}}
						A(\theta,1-\lambda \rho,P_{\bx \by})
				\right]
				}
\nonumber\\
& &										
			e^{N
				\left[
					\lambda \rho {\xi E_{r}^{*}(\theta')}
					-\rho \cdot
					\min_{P_{\bxtag|\by}}
					 A(\theta',\lambda,P_{\bxtag \by})	
				\right]
				}
	\Biggr\}
\nonumber\\
\label{E[S_N] upper bound_8_2}
& = &
\left| \Theta_N \right|^2
				\sum\limits_{T_{\by}}
				\max_{\theta \in \Theta_N}
				\max_{\theta'\in \Theta_N}
					\min_{\stackrel{0\leq\rho\leq1}{0\leq\lambda\leq 1/\rho}}
\nonumber\\
& &										
	\Biggl\{					
			\exp\Bigl\{
				N
				\bigl[
			 		\rho R
					+H_{\by}(Y)
					+(1- \lambda \rho) {\xi E_{r}^{*}(\theta)}
					-
					\min_{P_{\bx|\by}}
							A(\theta,1-\lambda \rho,P_{\bx \by})	
\nonumber\\
& &										
					+\lambda \rho {\xi E_{r}^{*}(\theta')}
					-\rho \cdot
					\min_{P_{\bxtag|\by}}
 							A(\theta',\lambda,P_{\bxtag \by})	
				\bigr]
				\Bigr\}
	\Biggr\}
\nonumber\\
\label{E[S_N] upper bound_8_3}
& \stackrel{(b)}{\leq} &
\left| \Theta_N \right|^2
(N+1)^{|\calY|}
			\max_{P_{\by}}												
				\max_{\theta \in \Theta_N}
				\max_{\theta'\in \Theta_N}
					\min_{\stackrel{0\leq\rho\leq1}{0\leq\lambda\leq 1/\rho}}
\nonumber\\
& &										
							\biggl\{					
										\exp
											\Bigl\{N
												\bigl[
											 		\rho R
													+H_{\by}(Y)
													+(1- \lambda \rho) {\xi E_{r}^{*}(\theta)}
													-
													\min_{P_{\bx|\by}}
															A(\theta,1-\lambda \rho,P_{\bx \by})	
\nonumber\\
& &										
													+\lambda \rho {\xi E_{r}^{*}(\theta')}
													-\rho \cdot 
													\min_{P_{\bxtag|\by}}
															A(\theta',\lambda,P_{\bxtag \by})	
												\bigr]
											\Bigr\}								
						\biggr\}
\nonumber\\						
\label{E[S_N] upper bound_9}
& = &
\left| \Theta_N \right|^2
(N+1)^{|\calY|}
\cdot
\exp
\biggl\{
		N \cdot
		\max_{P_{\by}}												
		\max_{\theta \in \Theta_N}
		\max_{\theta'\in \Theta_N}
		\min_{\stackrel{0\leq\rho\leq1}{0\leq\lambda\leq 1/\rho}}
		\max_{P_{\bx|\by}}
		\max_{P_{\bxtag|\by}}
\nonumber\\
& &										
												\bigl[
											 		\rho R
													+H_{\by}(Y)
													+(1- \lambda \rho) {\xi E_{r}^{*}(\theta)}
													-A(\theta,1-\lambda \rho,P_{\bx \by})	
\nonumber\\
& &										
													+\lambda \rho {\xi E_{r}^{*}(\theta')}
													-\rho \cdot
															A(\theta',\lambda,P_{\bxtag \by})	
												\bigr]
\biggr\},
\end{eqnarray}
where in ($a$) we upper bound $|T_{\by}|$ by 
$e^{N\cdot H_{\by}(Y)}$, and in ($b$) we upper bound the summation of the functional over $T_{\by}$ by the product of the maximal value (achieved by a specific distribution $P_{\by}$) with $(N+1)^{|\calY|}$, which is an upper bound to the number of type classes $\left\{T_{\by}\right\}$.

As explained earlier, we seek the maximal $\xi$ such that $\overline{S}_N$ grows sub--exponentially with $N$. To this end, we can ignore the factor 
$
\left| \Theta_N \right|^2
(N+1)^{|\calY|}
$
in (\ref{E[S_N] upper bound_9}), as it grows polynomially with $N$. Moreover, as mentioned in Section V, the optimizations can be conducted over continuous distributions and over the entire parameter space, $\Theta$. Thus, a maximal $\xi$ is sought, such that (using (\ref{B_definition})):
\begin{equation}
\label{optimal_xi}
				\max_{P_{Y}}												
				\max_{\theta,\theta' \in \Theta }
				\min_{\stackrel{0\leq\rho\leq1}{0\leq\lambda\leq 1/\rho}}
				\max_{P_{X|Y}}
				\max_{P_{X'|Y}}
				\Bigl[ 
						\rho R
						+(1- \lambda \rho) {\xi E_{r}^{*}(\theta)}
						+\lambda \rho {\xi E_{r}^{*}(\theta')}
						-	B(\theta,\theta',P_{Y},P_{X|Y},P_{X'|Y},\lambda,\rho)
				\Bigr]											 				
	\leq 0.
\end{equation}
An equivalent condition to (\ref{optimal_xi}) is\\
$\forall{\: P_{Y}},\: \forall{\: \theta,\theta' \in \Theta}$, $\exists{\: 0\leq\rho\leq 1,  0\leq\lambda\leq 1/\rho}$:\ \ \ \
$
\forall{\:
P_{X|Y}
},\: 
\forall{\: 
P_{X'|Y}
}
$
$$
	\rho R
	+(1- \lambda \rho) {\xi E_{r}^{*}(\theta)}
	+\lambda \rho {\xi E_{r}^{*}(\theta')}
	-B(\theta,\theta',P_{Y},P_{X|Y},P_{X'|Y},\lambda,\rho)
	\leq 0
$$
or,\\
$\forall{\: P_{Y}},\: \forall{\: \theta,\theta' \in \Theta}$, $\exists{\: 0\leq\rho\leq 1,  0\leq\lambda\leq 1/\rho}$:\ \ \ \ 
$
\forall{\:
P_{X|Y}
},\: 
\forall{\: 
P_{X'|Y}
}
$
$$
\xi \leq
		\frac{
							B(\theta,\theta',P_{Y},P_{X|Y},P_{X'|Y},\lambda,\rho)
							-\rho R
		}
		{	
			(1- \lambda \rho) \cdot E_{r}^{*}(\theta)
			+\lambda \rho \cdot E_{r}^{*}(\theta')
		}.
$$
Consequently, for ensembles of linear codes and BIOS channels, the lower bound to $\xi^{*}$ is the same as in (\ref{xi star - lower bound}), with a uniform i.i.d. random coding distribution, $Q^{*}=\left\{\frac{1}{2},\frac{1}{2}\right\}$.

\subsubsection*{Systematic Linear Codes}

A similar technique will be used now to achieve identical results for the ensemble of systematic linear codes.

By averaging $S_N$ over this ensemble:
\begin{eqnarray}
\label{SLC E[S_N] upper bound_1}
\overline{S}_N 
&\stackrel{(a)}{=}&
		2^{-K\left(N-K\right)}
			2^{-N}
				\sum\limits_{\tilde{\bG}}			
				\sum\limits_{\bvzero}
				S_N\left(\bvzero,\bG\right)
\nonumber\\
\label{SLC E[S_N] upper bound_2}
&\leq&
		2^{-K\left(N-K\right)}
			2^{-N}
				\sum\limits_{\tilde{\bG}}			
				\sum\limits_{\bvzero}
					\sum_{\by \in Y^N}
								\max_{\theta \in \Theta_N}
									\max_{\theta' \in \Theta_N}
										\min_{\stackrel{\rho\geq0}{0\leq\lambda\leq 1/\rho}}
\nonumber\\					
& &											
												\Biggl\{					
													{e^{N ({1-\lambda\rho}) {f_{\theta}(\bvzero,\by)}	}}
													\Biggl[
														\sum\limits_{m=1}^{M-1}
																	{e^{N \lambda f_{\theta'}
																	\left(
																				\bvm,\by
																	\right)}}
													\Biggr]^\rho
												\Biggr\}
\nonumber\\
\label{SLC E[S_N] upper bound_8}
&\stackrel{(b)}{\leq}&
\left| \Theta_N \right|^2
			\sum_{\by \in Y^N}
					\max_{\theta \in \Theta_N}
						\max_{\theta' \in \Theta_N}
						\min_{\stackrel{0\leq\rho\leq1}{0\leq\lambda\leq 1/\rho}}
\nonumber\\					
& &											
								\biggl\{					
									2^{-N}
									\sum\limits_{\bvzero}
										{e^{N ({1-\lambda\rho}) {f_{\theta}(\bvzero,\by)}	}}
										\Biggl[
											2^{-K(N-K)}
											\sum\limits_{m=1}^{M-1}
													\sum\limits_{\tilde{\bG}}										
															{e^{N \lambda f_{\theta'}
																\left(
																		\bvm,\by
																\right)}}
										\Biggr]^\rho
								\biggr\}
\nonumber\\
\label{SLC E[S_N] upper bound_11}
&\stackrel{(c)}{=}&
\left| \Theta_N \right|^2
			\sum_{\by \in Y^N}
					\max_{\theta \in \Theta_N}
						\max_{\theta' \in \Theta_N}
						\min_{\stackrel{0\leq\rho\leq1}{0\leq\lambda\leq 1/\rho}}
								\biggl\{					
									2^{-N}
									\sum\limits_{\bvzero}									
										{e^{N ({1-\lambda\rho}) {f_{\theta}(\bvzero,\by)}	}}
										\Biggl[
											2^{-(N-K)}
													\sum\limits_{\bv}
														{e^{N \lambda f_{\theta'}
															\left(
																{\bv}
																,\by
															\right)}}
										\Biggr]^\rho
								\biggr\}
\nonumber\\
\label{SLC E[S_N] upper bound_12}
& \stackrel{(d)}{=}&
\left| \Theta_N \right|^2
			\sum_{\by \in Y^N}
					\max_{\theta \in \Theta_N}
						\max_{\theta' \in \Theta_N}
						\min_{\stackrel{0\leq\rho\leq1}{0\leq\lambda\leq 1/\rho}}
								\biggl\{					
									2^{-N}
									\sum\limits_{\bvzero}
										{e^{N ({1-\lambda\rho}) {f_{\theta}(\bvzero,\by)}	}}
										\Biggl[
											M2^{-N}
													\sum\limits_{\bv}
														{e^{N \lambda f_{\theta'}
															\left(
																{\bv}
																,\by
															\right)}}
										\Biggr]^\rho
								\biggr\}
\nonumber\\
\label{SLC E[S_N] upper bound_13}
&=&
\left| \Theta_N \right|^2
			\sum_{\by \in Y^N}
					\max_{\theta \in \Theta_N}
						\max_{\theta' \in \Theta_N}
						\min_{\stackrel{0\leq\rho\leq1}{0\leq\lambda\leq 1/\rho}}
								\biggl\{					
									M^{\rho}
\nonumber\\					
& &											
									\Bigl[
										2^{-N}
										\sum\limits_{\bv}
											{e^{N ({1-\lambda\rho}) {f_{\theta}(\bv,\by)} }}
									\Bigr]
										\Bigl[
											2^{-N}
													\sum\limits_{\bvtag}
														{e^{N \lambda f_{\theta'}
															\left(
																{\bvtag}
																,\by
															\right)}}
										\Bigr]^\rho.
								\biggr\}
\nonumber\\
\label{SLC E[S_N] upper bound_14}
\end{eqnarray}
The equality in ($a$) is obtained by averaging over $2^{K\left(N-K\right)}$ and $2^{N}$ equiprobable values of $\bvzero$ and $\tilde{\bG}$ (the non--systematic part of $\bG$), respectively.
($b$) is obtained by taking identical steps as done for ensemble of linear codes in the previous subsection (see the inequalities between (\ref{LC E[S_N] upper bound_2_A}) and (\ref{LC E[S_N] upper bound_3_2_A})).
In ($c$), we used the following equivalence for the two inner summations:
\begin{eqnarray}
\label{SLC E[S_N] inner term equivalence}
 \sum\limits_{m=1}^{M-1}
 \sum\limits_{\tilde{\bG}}										
	{e^{N \lambda f_{\theta'}
			\left(
				\bvm,\by
			\right)}}
=
 2^{(K-1)(N-K)}
 \sum\limits_{\bv}
  	{e^{N \lambda f_{\theta'}
			\left(
				\bv
				,\by
		\right)}}.
\end{eqnarray}
This equivalence is proved in Section A.6 of the Appendix. In ($d$), we used the equality $M=2^K$.

Finally, the upper bound to $\overline{S}_N$ achieved in (\ref{SLC E[S_N] upper bound_14}) is identical to the one related to ensembles of linear codes (see (\ref{LC E[S_N] upper bound_13})), and therefore the final lower bound to $\xi^{*}$ for the case of systematic linear codes is also identical to (\ref{xi star - lower bound}) with uniform i.i.d. random coding distribution, $Q^{*}=\left\{\frac{1}{2},\frac{1}{2}\right\}$.

\subsection*{A.2 Proof of eq. (\ref{inner_term_1_optimizations_interchange-3}) and eq. (\ref{LC S_N inner optimization equivalence})}

Let $\theta^{*}$ maximize $f_{\theta}(\bvm,\by)$, and let $F(\lambda,\rho)$ be a nonnegative function. Then,

\begin{lefteqn}
{
				\min_{\stackrel{\rho\geq0}{0\leq\lambda\leq 1/\rho}}
				\max_{\theta \in \Theta_N}
							\Biggl\{					
										{e^{N({1-\lambda\rho})
											{f_{\theta}(\bvm,\by)}}}
								\cdot
								F(\lambda,\rho)
							\Biggr\}
=
\nonumber
}
\end{lefteqn}
\begin{eqnarray}
& = &
				\min_{\stackrel{\rho\geq0}{0\leq\lambda\leq 1/\rho}}
							\Biggl\{					
										{e^{N({1-\lambda\rho})
											{f_{\theta^{*}}(\bvm,\by)}}}
								\cdot
								F(\lambda,\rho)
							\Biggr\}
\nonumber\\
& \stackrel{(a)}{\leq} &
						\max_{\theta\in \Theta_N}
						\min_{\stackrel{\rho\geq0}{0\leq\lambda\leq 1/\rho}}
							\Biggl\{					
										{e^{N({1-\lambda\rho})
											{f_{\theta}(\bvm,\by)}}}
								\cdot
								F(\lambda,\rho)
							\Biggr\}
\nonumber\\
& \leq &
				\min_{\stackrel{\rho\geq0}{0\leq\lambda\leq 1/\rho}}
				\max_{\theta \in \Theta_N}
							\Biggl\{					
										{e^{N({1-\lambda\rho})
											{f_{\theta}(\bvm,\by)}}}
								\cdot
								F(\lambda,\rho)
							\Biggr\},
\end{eqnarray}
where (a) is true since the value of the function for a specific $\theta^{*}$ in $\Theta_N$ is always upper--bounded by the maximization of the function over $\theta \in \Theta_N$. Thus, all inequalities must be achieved with equalities.

\subsection*{A.3 Proof of eq. (\ref{E upper bound})}

For $\alpha\in\Re$ and $\by \in {\cal Y}^N$, we exponentially evaluate 
$
E
		\bigl[
					{e^{N{\alpha}{f_{\theta}(\bx,\by)}
						}}
		\bigr]
$
, where the average is calculated over the ensemble of random coding distribution of the form:
$
Q_N(\bx)=\frac{Q_N(T_{\bx})}{\left|T_{\bx}\right|},
$
(as described in (\ref{E upper bound})):
\begin{eqnarray}
\label{TBD_MOT_UB_1}
E
		\bigl[
					{e^{N{\alpha}{f_{\theta}(\bX,\by)}
						}}
		\bigr]
&=&
\sum\limits_{\bx\in\calX^N}
					Q_N(\bx)
					{e^{N{\alpha}{f_{\theta}(\bx,\by)}
						}}
\nonumber\\
\label{TBD_MOT_UB_2}
&\stackrel{(a)}{=}&
\sum\limits_{T_{\bx|\by}\subset\calX^N}
					\left|T_{\bx|\by}\right|
					Q_N(\bx)
					{e^{N{\alpha}{f_{\theta}(\bx,\by)}
						}}
\nonumber\\
\label{TBD_MOT_UB_2.5}
&=&
\sum\limits_{T_{\bx|\by}\subset\calX^N}
					\left|T_{\bx|\by}\right|
					\frac{e^{-N\Delta^{*}_N(P_{\bx})}}{|T_{\bx}|}
					{e^{N{\alpha}{f_{\theta}(\bx,\by)}
						}}
\nonumber\\
\label{TBD_MOT_UB_2.6}
&\stackrel{(b)}{=}&
\sum\limits_{T_{\bx|\by}\subset\calX^N}
					\left|T_{\bx|\by}\right|
					\frac{e^{-N(\Delta^{*}(P_{\bx})-\tilde{\epsilon}_N)}}{|T_{\bx}|}
					{e^{N{\alpha}{f_{\theta}(\bx,\by)}
						}},
\end{eqnarray}						
where $\tilde{\epsilon}_N\rightarrow 0$ as $N\rightarrow \infty$ independently of $P_{\bx}$.\\			
We should note that ($a$) is true since $f_{\theta}(\bx,\by)$ depends on $\bx$ and $\by$ only via their joint empirical distribution and the summation can be conducted over types instead, and since the average is calculated for a given $\by$, we sum over $T_{\bx|\by}$.
In ($b$) we used the convergence assumption for the random coding distributions withing the class $\calQ$.\\
Thus, we continue to evaluate
$
E
		\bigl[
					{e^{N{\alpha}{f_{\theta}(\bx,\by)}
						}}
		\bigr]
$
as follows:
\begin{eqnarray}
\label{TBD_MOT_UB_3}
E
		\bigl[
					{e^{N{\alpha}{f_{\theta}(\bx,\by)}
						}}
		\bigr]
&\stackrel{(a)}{\stackrel{\cdot}{=}}&
\sum\limits_{T_{\bx|\by}\subset\calX^N}
					\exp\Bigl\{N
								\bigl[
											-I_{\bx\by}(X;Y)
											-\Delta^{*}(P_{\bx})
											+{\alpha}f_{\theta}(\bx,\by)
								\bigr]											 
					\Bigr\}
\nonumber\\
\label{TBD_MOT_UB_5}
&\stackrel{(b)}{\stackrel{\cdot}{=}}&
\exp\Bigl\{
	N\cdot 
	\max_{P_{\bx|\by}}
			\bigl[
						 -I_{\bx\by}(X;Y)
							-\Delta^{*}(P_{\bx})
						 +{\alpha}f_{\theta}(\bx,\by)
			\bigr]
		\Bigr\}
\nonumber\\
\label{TBD_MOT_UB_6}
&\stackrel{(c)}{=}&
	\exp\Bigl\{N\cdot 
		\max_{P_{\bx|\by}}
				\bigl[
						 	-I_{\bx\by}(X;Y)
							-\Delta^{*}(P_{\bx})
\nonumber\\
& &											
\ \ \ \
						 	+{\alpha}\hat{E}_{\bx\by}
												\ln{
													P_\theta({Y}|{X})
													}
							+{\alpha}{\xi E_{r}^{*}(\theta)}
					\bigr]
\Bigr\}
\nonumber\\
\label{TBD_MOT_UB_7}
&=&
\exp\Bigl\{
	N\bigl[ 
						 	{\alpha}{\xi E_{r}^{*}(\theta)}
						 	-\min_{P_{\bx|\by}}
						 	\bigl\{
						 		I_{\bx\by}(X;Y)
\nonumber\\					
& &											
\ \
						+\Delta^{*}(P_{\bx})
				 		-{\alpha}\hat{E}_{\bx\by}
												\ln{
													P_\theta({Y}|{X})
													}
							\bigr\}
				\bigr]											 				
		\Bigr\}
\nonumber\\
\label{TBD_MOT_UB_8}
&=&
e^{
	N\bigl[ 
			 	{\alpha}{\xi E_{r}^{*}(\theta)}
			 	-\min_{P_{\bx|\by}}
						 		A(\theta,\alpha,P_{\bx \by})
   \bigr]
	},
\end{eqnarray}
where in ($a$), we used the facts that  
$
\left|
		T_{\bx|\by}
\right|
\stackrel{\cdot}{=}
e^{N\cdot H_{\bx\by}(X|Y)}			
$
and
$
\left|
		T_{\bx}
\right|
\stackrel{\cdot}{=}
e^{N\cdot H_{\bx}(X)}
$.
($b$) is true since the summation of the functional over $T_{\bx|\by}\subset\calX^N$ is lower bounded by its maximal value (achieved by a specific distribution $P_{\bx|\by}$), and upper bounded by the product of its maximal value with
$
	(N+1)^{|\calX||\calY|}
$.
In ($c$), we expressed the minimax metric in terms of the joint empirical distribution as described in (\ref{f dep on x n y-5}).

\subsection*{A.4 Proof of Lemma 2}

In this section, we prove that when a linear code is used for a BIOS channel and the minimax decision rule is used (denoted by $\Omega$), the error probability for the $m$-th message (of length N), $\bvm = (v_{m0},\ldots,v_{m(N-1)})$, is the same for all $m$, that is,
\begin{eqnarray}
\label{error expo indep of m}
P_{E_m}\left(
					\Omega|\theta
			 \right)
=P_E\left(
				\Omega|\theta
			\right)
\ \ \ \ \ for \ \ 0\leq m\leq M-1.
\end{eqnarray}
%
%
%
%
Considering a binary input channel, we denote the channel crossover probabilities for a single letter as \ \ \ 
$
P_{\theta}(y|v=0)\stackrel{\Delta}{=} P_{\theta,0}(y)\ \ 
$
and  \ \ \ 
$
P_{\theta}(y|v=1)\stackrel{\Delta}{=} P_{\theta,1}(y)
$.\\
If the channel is also output symmetric then,
\begin{eqnarray}
P_{\theta,1}(y) = P_{\theta,0}(-y), \ \ \ \forall y \in {\cal Y}
\nonumber
\end{eqnarray}
The error probability for the $m$-th message using minimax decoding is:
\begin{eqnarray}
\label{Appendix - error expo-1}
P_{E_m}\left(
					\Omega|\theta
			 \right)
& = & 
\sum_{\by\in{\Lambda_m}^c} P_{\theta}(\by|\bvm) \nonumber\\
\label{Appendix - error expo-2}
& = & 
\sum_{\by\in{\Lambda_m}^c}
			\prod_{n:v_{mn}=0} P_{\theta,0}(y_{n})
			\prod_{n:v_{mn}=1} P_{\theta,1}(y_{n}) \nonumber\\
\label{Appendix - error expo-3}
& = & 
\sum_{\by\in{\Lambda_m}^c}
			\prod_{n:v_{mn}=0} P_{\theta,0}(y_{n})
			\prod_{n:v_{mn}=1} P_{\theta,0}(-y_{n}),
\nonumber\\
\end{eqnarray}
where
\begin{eqnarray}
\label{Appendix-the decision area complement_1}
{\Lambda_m}^c
&=&
	\Biggl\{
	\by:
			\max_{\theta'}
					\left\{\frac{1}{N}
							\ln{P_{\theta'}(\by|\bvmtag)}+
							{\xi E_{r}^{*}({\theta'})}
					\right\}
			\geq
		\max_{\theta''}
					\left\{\frac{1}{N}
							\ln{P_{\theta''}(\by|\bvm)}+
							{\xi E_{r}^{*}({\theta''})}
					\right\}			
			,
\nonumber\\
& &			
\ \ \ \ \ \ 
		 for\ some\ {m'}\neq m
		\Biggr\} \nonumber\\
\label{Appendix-the decision area complement_2}		
&=&
	\Biggl\{
	\by:
			\max_{\theta'}
					\left\{
							\sum_{n=0}^{N-1}\ln{P_{\theta'}\left(y_{n}|v_{m'n}\right)}+
							{N\xi E_{r}^{*}({\theta'})}
					\right\}
			\geq
\nonumber\\
& &			
\ \ \ \ \ \ 
		\max_{\theta''}
					\left\{
							\sum_{n=0}^{N-1}\ln{P_{\theta''}\left(y_{n}|v_{mn}\right)}+
							{N\xi E_{r}^{*}({\theta''})}
					\right\}			
			,
\ \ \ \ 
			for\ some\ {m'}\neq m
		\Biggr\} \nonumber\\
\label{Appendix-the decision area complement_3}		
&=&	\Biggl\{
		\by:
			\max_{\theta'}
				\Bigl\{
					\sum_{\stackrel{t:\ v_{mt}=0}{v_{m't}=0}}\ln P_{{\theta'},0}(y_{t})+
					\sum_{\stackrel{t:\ v_{mt}=0}{v_{m't}=1}}\ln P_{{\theta'},1}(y_{t})+
					\sum_{\stackrel{t:\ v_{mt}=1}{v_{m't}=0}}\ln P_{{\theta'},0}(y_{t})+
\nonumber\\
& &			
\ \ \ \ \ \ \ \ \ \ \ \ 
					\sum_{\stackrel{t:\ v_{mt}=1}{v_{m't}=1}}\ln P_{{\theta'},1}(y_{t})+
					N\xi E_{r}^{*}(\theta')
				\Bigr\}
			\geq \nonumber\\
& &
\ \ \ \ \ \ 
	\max_{\theta''}
				\Bigl\{
					\sum_{\stackrel{t:\ v_{mt}=0}{v_{m't}=0}}\ln P_{{\theta''},0}(y_{t})+
					\sum_{\stackrel{t:\ v_{mt}=0}{v_{m't}=1}}\ln P_{{\theta''},0}(y_{t})+
					\sum_{\stackrel{t:\ v_{mt}=1}{v_{m't}=0}}\ln P_{{\theta''},1}(y_{t})+
\nonumber\\
& &			
\ \ \ \ \ \ \ \ \ \ \ \ 
					\sum_{\stackrel{t:\ v_{mt}=1}{v_{m't}=1}}\ln P_{{\theta''},1}(y_{t})+
					N\xi E_{r}^{*}(\theta'')
				\Bigr\}
			,
\ \ \ \ \ \ 
			 for\ some\ {m'}\neq m
	\Biggr\}\nonumber\\
\label{Appendix-the decision area complement_4}	
&=&	\Biggl\{
		\by:
			\max_{\theta'}
				\Bigl\{
					\sum_{\stackrel{t:\ v_{mt}=0}{v_{m't}=0}}\ln P_{{\theta'},0}(y_{t})+
					\sum_{\stackrel{t:\ v_{mt}=0}{v_{m't}=1}}\ln P_{{\theta'},0}(-y_{t})+
					\sum_{\stackrel{t:\ v_{mt}=1}{v_{m't}=0}}\ln P_{{\theta'},0}(y_{t})+
\nonumber\\
& &			
\ \ \ \ \ \ \ \ \ \ \ \ 
					\sum_{\stackrel{t:\ v_{mt}=1}{v_{m't}=1}}\ln P_{{\theta'},0}(-y_{t})+
					N\xi E_{r}^{*}(\theta')
				\Bigr\}
			\geq 
\nonumber\\
& &			
\ \ \ \ \ \ 
	\max_{\theta''}
				\Bigl\{
					\sum_{\stackrel{t:\ v_{mt}=0}{v_{m't}=0}}\ln P_{{\theta''},0}(y_{t})+
					\sum_{\stackrel{t:\ v_{mt}=0}{v_{m't}=1}}\ln P_{{\theta''},0}(y_{t})+
					\sum_{\stackrel{t:\ v_{mt}=1}{v_{m't}=0}}\ln P_{{\theta''},0}(-y_{t})+
\nonumber\\
& &			
\ \ \ \ \ \ \ \ \ \ \ \ 
					\sum_{\stackrel{t:\ v_{mt}=1}{v_{m't}=1}}\ln P_{{\theta''},0}(-y_{t})+
					N\xi E_{r}^{*}(\theta'')
				\Bigr\}
			,
\ \ \ \ \ \ 
	 for\ some\ {m'}\neq m
	\Biggr\}.
\end{eqnarray}
Using the following transformation to dummy variables
$$z_{n}=\Biggl\{
\stackrel{y_{n}, \ \ \ \ \forall n\ : v_{mn}=0}
	{-y_{n},\ \ \ \  \forall n\ : v_{mn}=1}
$$
we get that
\begin{eqnarray}
\label{Appendix - error expo updated-1}
P_{E_m}\left(
					f|\theta
			 \right)
& = & 
\sum_{\bz\in{\Lambda_m}^c}
			\prod_{n:v_{mn}=0} P_{\theta,0}(z_{n})
			\prod_{n:v_{mn}=1} P_{\theta,0}(z_{n})
\nonumber\\
\label{Appendix - error expo updated-2}
& = & 
\sum_{\bz\in{\Lambda_m}^c}
			\prod_{n=0}^{N-1} P_{\theta,0}(z_{n}),
\end{eqnarray}
where
\begin{eqnarray}
\label{Appendix-the decision area complement-updated_1}
{\Lambda_m}^c 
&=&
	\Biggl\{
		\bz:
			\max_{\theta'}
				\Bigl\{
					\sum_{\stackrel{t:\ v_{mt}=0}{v_{m't}=0}}\ln P_{{\theta'},0}(z_{t})+
					\sum_{\stackrel{t:\ v_{mt}=0}{v_{m't}=1}}\ln P_{{\theta'},0}(-z_{t})+
					\sum_{\stackrel{t:\ v_{mt}=1}{v_{m't}=0}}\ln P_{{\theta'},0}(-z_{t})+
\nonumber\\
& &			
\ \ \ \ \ \ \ \ \ \ \ \ 
					\sum_{\stackrel{t:\ v_{mt}=1}{v_{m't}=1}}\ln P_{{\theta'},0}(z_{t})+
					N\xi E_{r}^{*}(\theta')
				\Bigr\}
			\geq
\nonumber\\
& &			
\ \ \ \ \ \ 
	\max_{\theta''}
				\Bigl\{
					\sum_{\stackrel{t:\ v_{mt}=0}{v_{m't}=0}}\ln P_{{\theta''},0}(z_{t})+
					\sum_{\stackrel{t:\ v_{mt}=0}{v_{m't}=1}}\ln P_{{\theta''},0}(z_{t})+
					\sum_{\stackrel{t:\ v_{mt}=1}{v_{m't}=0}}\ln P_{{\theta''},0}(z_{t})+
\nonumber\\
& &			
\ \ \ \ \ \ \ \ \ \ \ \ 
					\sum_{\stackrel{t:\ v_{mt}=1}{v_{m't}=1}}\ln P_{{\theta''},0}(z_{t})+
					N\xi E_{r}^{*}(\theta'')
				\Bigr\}
,
\ \ \ \ \ \ 
		 for\ some\ {m'}\neq m
	\Biggr\} \nonumber\\
\label{Appendix-the decision area complement-updated_2}	
&=&
	\Biggl\{
		\bz:
			\max_{\theta'}
				\Bigl\{
					\sum_{p:{v_{mp}=v_{m'p}}}\ln P_{{\theta'},0}(z_{p})+
					\sum_{q:{v_{mq}\neq v_{m'q}}}\ln P_{{\theta'},0}(-z_{q})+
					N\xi E_{r}^{*}(\theta')
				\Bigr\}
			\geq
\nonumber\\
& &			
\ \ \ \ \ \ 
	\max_{\theta''}
				\Bigl\{
					\sum_{p:{v_{mp}=v_{m'p}}}\ln P_{{\theta''},0}(z_{p})+
					\sum_{q:{v_{mq}\neq v_{m'q}}}\ln P_{{\theta''},0}(z_{q})+
					N\xi E_{r}^{*}(\theta'')
				\Bigr\}
			,
\nonumber\\
& &			
\ \ \ \ \ \ 
		 for\ some\ {m'}\neq m
	\Biggr\}.
\end{eqnarray}
Now, on the one hand, (\ref{Appendix - error expo updated-2}) and (\ref{Appendix-the decision area complement-updated_2}) describe $P_{E_m}(f|\theta)$ and ${\Lambda_m}^c$, respectively, for each $0 \leq m \leq M-1$. On the other hand, we should note that the terms for $P_{E_0}(f|\theta)$ and ${\Lambda_0}^c$ (describing the case where $\bvzero=\bzero$ is transmitted) are obtained by assigning $m=0$ in (\ref{Appendix - error expo-3}) and (\ref{Appendix-the decision area complement_4}). By doing that, the result terms coincide with (\ref{Appendix - error expo updated-2}) and (\ref{Appendix-the decision area complement-updated_2}), respectively (which, as mentioned before, correspond to the $m$-th message).
This observation completes the proof.

\subsection*{A.5 Proof of eq. (\ref{LC E[S_N] inner term equivalence})}

First, by the way of constructing the linear code, we know that:
\begin{eqnarray}
\label{APP_LC_inner term equivalence - 1}
\bvmtag = \bumtag \bG\oplus\bvzero, \ \ \ \ \ \forall \ 0 \leq m' \leq M-1
\end{eqnarray}
Since $1 \leq m' \leq M-1$ implies $\bumtag\neq\bzero$, then for each information vector in this set there is at least one index $i$ for which ${u}_{m'i}=1$.
Consequently, the construction of each code vector $\bvmtag$, $1 \leq m' \leq M-1$, can be written in the following way:
\begin{eqnarray}
\label{LC E[S_N] inner term 2}
\bvmtag 
=
\bumtag \bG\oplus\bvzero
=
\bgi\oplus
\Bigl[
	\sum\limits_{j\neq i}u_{m'j}\bgj
\Bigr]\oplus
\bvzero,
\nonumber
\end{eqnarray}
where $\bgi$ stands for the $i$-th row in $\bG$.\\
Therefore:
\begin{eqnarray}
\label{LC E[S_N] inner term 3}
\sum\limits_{m'=1}^{M-1}
\sum\limits_{\bG}										
	{e^
		{
			N \lambda f_{\theta'}
			\left(
					\bvmtag,\by
			\right)
		}
	}
&=&
\sum\limits_{m'=1}^{M-1}
\sum\limits_{\bG \backslash {\bgi}}
\sum\limits_{\bgi}
	\exp
		\Bigl\{
		N
		f_{\theta'}
			\bigl(
					\bgi\oplus
					\Bigl[ 
						\sum\limits_{j\neq i}u_{m'j}\bgj
					\Bigr]\oplus
					\bvzero
					,\by
			\bigr)\lambda
			\Bigr\}
\nonumber\\
\label{LC E[S_N] inner term 4}
& \stackrel{(a)}{=}&
\sum\limits_{m'=1}^{M-1}
\sum\limits_{\bG \backslash {\bgi}}
\sum\limits_{\bv}
	{e^{Nf_{\theta'}
		\left(
				\bv
				,\by
		\right)\lambda}}
\nonumber\\
\label{LC E[S_N] inner term 5}
&=&
\sum\limits_{m'=1}^{M-1}
2^{(K-1)N}
\sum\limits_{\bv}
	{e^{Nf_{\theta'}
		\left(
				\bv
				,\by
		\right)\lambda}}
\nonumber\\
\label{LC E[S_N] inner term 6}
\nonumber\\
&=&
(M-1)
2^{(K-1)N}
\sum\limits_{\bv}
	{e^{Nf_{\theta'}
		\left(
				\bv
				,\by
		\right)\lambda}},
\nonumber\\
\end{eqnarray}
where ($a$) is true since for fixed values of $m'$, $\bG \backslash {\bgi}$ (in the outer summations) and $\bvzero$, the row vector, which is denoted by $\bv$, is fixed, causing $\bgi$ to sum up over all the binary vectors of length $N$.

\subsection*{A.6 Proof of eq. (\ref{SLC E[S_N] inner term equivalence})}

In this section, we prove the equality, which is given in (\ref{SLC E[S_N] inner term equivalence}), and used in (\ref{SLC E[S_N] upper bound_13}).\\
First, by the way of constructing a systematic linear code:
\begin{eqnarray}
\label{APP_SLC_inner term equivalence - 1}
\bvmtag & = & \bumtag \bG\oplus\bvzero
\nonumber\\
&=&
		\Bigl[ 
				\bumtag ;
				\sum\limits_{i=1}^{K}u_{m'i}\bgitilde
		\Bigr]
		\oplus
		\bvzero
\nonumber\\
&=&
		\Bigl[ 
				\bumtag ; \overbrace{0\ldots0}^{N-K}
		\Bigr]
		\oplus
		\Bigl[ 
			\overbrace{0\ldots0}^{K} ; \sum\limits_{i=1}^{K}u_{m'i}\bgitilde
		\Bigr]
		\oplus
		\bvzero
 ,
 \ \ \ \ \ 
 \forall \ 0 \leq m' \leq M-1,
\end{eqnarray}
where $\bgitilde$ stands for the $i$'th row in $\tilde{\bG}$ (the non-systematic part of $\bG$).\\
We observe that for $1 \leq m' \leq M-1$, $\bumtag\neq\bzero$. Thus, for each information vector in this set there's at least one index $i$ for which ${u}_{m'i}=1$.
Consequently, the construction of each code vector $\bvmtag$, $1 \leq m' \leq M-1$, can be written in the following way:

\begin{eqnarray}
\label{SLC E[S_N] inner term 2}
\bvmtag
=
		\Bigl[ 
				\bumtag ; \bgitilde
		\Bigr]
		\oplus
		\Bigl[ 
			\overbrace{0\ldots0}^{K} ; \sum\limits_{j\neq i}u_{m'j}\bgjtilde
		\Bigr]
		\oplus
		\bvzero.
\nonumber
\end{eqnarray}
Therefore:
\begin{eqnarray}
\label{SLC E[S_N] inner term 3}
\sum\limits_{m'=1}^{M-1}
	\sum\limits_{\tilde{\bG}}										
		{e^{Nf_{\theta'}
			\left(
					\bvmtag,\by
			\right)\lambda}}
&=&
\sum\limits_{m'=1}^{M-1}
\sum\limits_{\tilde{\bG} \backslash {\tilde{\bg}_i}}
\sum\limits_{\tilde{\bg}_i}
	\exp\biggl\{
		N
		f_{\theta'}
		\Bigl(
				\left[ 
					\bumtag ; \bgitilde
				\right]
				\oplus
				\Bigl[ 
					0\ldots0 ; \sum\limits_{j\neq i}u_{m'j}\bgjtilde
				\Bigr]
				\oplus
				\bvzero
				,\by
		\Bigr)\lambda
		\biggr\}
\nonumber\\
& \stackrel{(a)}{=}&
\sum\limits_{\tilde{\bG}  \backslash {\tilde{\bg}_i}}
\sum\limits_{m'=1}^{M-1}
\sum\limits_{\tilde{\bg}_i}
	{e^{Nf_{\theta'}
		\Bigl(
				\left[ 
					\bumtag ; \tilde{\bg}_i
				\right]
				\oplus
				\bv
				,\by
		\Bigr)\lambda}}
\nonumber\\
& \stackrel{(b)}{\leq}&
\sum\limits_{\tilde{\bG} \backslash {\tilde{\bg}_i}}
\sum\limits_{m'=0}^{M-1}
\sum\limits_{\tilde{\bg}_i}
	{e^{Nf_{\theta'}
		\Bigl(
				\left[ 
					\bumtag ; \tilde{\bg}_i
				\right]
				\oplus
				\bv
				,\by
		\Bigr)\lambda}}
\nonumber\\
\label{SLC E[S_N] inner term 4}
& \stackrel{(c)}{=}&
\sum\limits_{\tilde{\bG} \backslash {\tilde{\bg}_i}}
\sum\limits_{\bv}
	{e^{Nf_{\theta'}
		\bigl(
				\bv
				,\by
		\bigr)\lambda}}
\nonumber\\
\label{SLC E[S_N] inner term 5}
&=&
2^{(K-1)(N-K)}
\sum\limits_{\bv}
	{e^{Nf_{\theta'}
		\bigl(
				\bv
				,\by
		\bigr)\lambda}},
\end{eqnarray}
where ($a$) is true since for fixed values of $m'$, $\tilde{\bG} \backslash {\bgitilde}$ (in the outer summations) and $\bvzero$, the row vector, which is denoted by $\bv$, is fixed.
In ($b$), $m'=0$ was added to the summation, and since the inner term in the summation is always non-negative the result cannot get smaller.
($c$) is true since for a fixed $\bv$, summing up over $0 \leq m' \leq M-1$ and $\bgitilde$ is equivalent to the summation over all the possibilities for a vector of length $N$.

\subsection*{A.7 Equivalence between decision rules - $\Omega$ and $\Lambda$}

In this section, we prove the equivalence between the minimax decision rule, $\Omega$, maximizing the metric $f(\bx,\by)$ (as defined in (\ref{explicit_decision_rule_2})), and a decision rule $\Lambda$, minimizing $\rho(\bx,\by)$ (as defined in (\ref{rho based decision rule})). We will prove that for a given output $\by \in {\cal Y}$, each $\bxone, \bxtwo \in {\cal X}$ satisfy:
\begin{eqnarray}
\label{f and rho equivalence 1}
	f(\bxone,\by) \geq
	f(\bxtwo,\by)
	\Longleftrightarrow
	\rho(\bxone,\by) \leq
	\rho(\bxtwo,\by).
\end{eqnarray}
First, we should note that
$
 	f(
						\bx,\by
	 )
$
satisfies:
\begin{eqnarray}
	f(
						\bx,\by
	 )
& = &
	\max_{0 \leq \theta \leq 1}
	f_\theta(
						\bx,\by
					)
\nonumber\\
 	& = &
	\max_{0 \leq \theta \leq 1} 	
	\Bigl\{
 	\frac{1}{N}
			\left[
					\ln{P_\theta(\by|	\bx)}+
					{N\xi E_{r}^{*}(\theta)}
			\right]
	\Bigr\}			
	\nonumber\\
	& \stackrel{(a)}{=} &
	\max_{0 \leq \theta \leq 1}
	\Bigl\{
 	\frac{1}{N}
			\Bigl[
				d(\bx,\by)\ln\theta
				+\left(
					N-d(\bx,\by)
				\right)
				\ln\left(1-\theta\right)
				+{N\xi E_{r}^{*}(\theta)}
			\Bigr]
	\Bigr\}			
	\nonumber\\
	& = &
	\max_{0 \leq \theta \leq 1}
	\Bigl\{
			\delta(\bx,\by)\ln\theta+
			\left(
				1-\delta(\bx,\by)
			\right)
			\ln\left(1-\theta\right)
			+{\xi E_{r}^{*}(\theta)}
	\Bigr\}
	\nonumber\\			
	& = &
	\max_{0 \leq \theta \leq 1}
	f_\theta\left(
						\delta(\bx,\by)
					\right),
	 \ \ \ 0 \leq \delta(\bx,\by) \leq 1
	\nonumber\\			
	& = &
	f\left(
						\delta(\bx,\by)
		\right),
	 \ \ \ 0 \leq \delta(\bx,\by) \leq 1.
\end{eqnarray}
In ($a$), we used the following representation for the BSC transition probability:
$$
P_{\theta}(\by|\bx) = \theta^{d(\bx,\by)}
																					\left(
																						1-\theta
																					\right)
																					^{N-d(\bx,\by)}.
$$
We conclude that the value of 
$
	f(
			\bx,\by
	)
$
is equal for all code vectors with the same (normalized) Hamming distance from $\by$, and therefore can be defined as 
$
	f(
						\delta(\bx,\by)
	 ),
	\ \ 0 \leq \delta(\bx,\by) \leq 1
$.

Next, we now prove that
$
f(
					\bx,\by
	)
$
 has the same value for a code vector $\bx$ and its complement, $\bolx$:
\begin{eqnarray}
	f(
					\bolx,\by
	)
 	& = &
 	f(
						\delta(\bolx,\by)
		)
	\nonumber\\
 	& = &
 	f(
						1-\delta(\bx,\by)
		)
	\nonumber\\
 	& = &
	\max_{0 \leq \theta \leq 1}
 	f_\theta(
 						1-\delta(\bx,\by)
 					)
	\nonumber\\
 	& = &
	\max_{0 \leq \theta \leq 1}
	\Bigl\{
		\left(
		1-\delta(\bx,\by)
		\right)
		\ln\theta
		+\delta(\bx,\by)
		\ln\left(1-\theta\right)+
		{\xi E_{r}^{*}(\theta)}
	\Bigr\}
	\nonumber\\
 	& \stackrel{(a)}{=} &
	\max_{0 \leq \tilde{\theta} \leq 1}
	\Bigl\{
		\left(
		1-\delta(\bx,\by)
		\right)
		\ln\left(1-\tilde{\theta}\right)
		+\delta(\bx,\by)
		\ln\tilde{\theta}+
		{\xi E_{r}^{*}(1-\tilde{\theta})}
	\Bigr\}
	\nonumber\\
 	& \stackrel{(b)}{=} &
	\max_{0 \leq \tilde{\theta} \leq 1}
	\Bigl\{
		\left(
		1-\delta(\bx,\by)
		\right)
		\ln\left(1-\tilde{\theta}\right)
		+\delta(\bx,\by)
		\ln\tilde{\theta}+
		{\xi E_{r}^{*}(\tilde{\theta})}
	\Bigr\}
	\nonumber\\ 	
 	& = &
	\max_{0 \leq \tilde{\theta} \leq 1}
	f_{\tilde{\theta}}(
						\delta(\bx,\by)
						)
	\nonumber\\ 	
 	& = &
 	f(
 				\delta(\bx,\by)
	)
	\nonumber\\ 	
 	& = &
 	f(
 				\bx,\by
	).
\end{eqnarray}
In ($a$), we changed the variable in the maximization, $\tilde{\theta} = 1- \theta$, and
($b$) is true since for the BSC model the ML error exponent, $E_{r}^{*}(\theta)$, is symmetric around $\theta=\frac{1}{2}$ (see (\ref{ML error_exponent - 2.5})).

Using the fact that both 
$
 	f(
 				\delta(\bx,\by)
	)
$
and
$
\rho(
 				\delta(\bx,\by)
	)
$
are equal for $\delta(\bx,\by)$ and $1-\delta(\bx,\by)$, it is sufficient to prove (\ref{f and rho equivalence 1}) for $\bxone$ and $\bxtwo$ satisfying
$
	\delta(\bxone,\by) \leq \frac{1}{2}
$
and
$
	\delta(\bxtwo,\by) \leq \frac{1}{2}
$ \ \ 
(and thus
$
\rho(
 				\bxone,\by
	)=
\delta(
				\bxone,\by
			)	
$
,
$
\rho(
 				\bxtwo,\by
	)=
\delta(
				\bxtwo,\by
			)	
$
).

In the rest of the proof, we will denote 
$
\delta(
				\bxone,\by
			)	
\stackrel{\Delta}{=}
\delta_{1}
$
,
$
\delta(
				\bxtwo,\by
			)	
\stackrel{\Delta}{=}
\delta_{2}			
$.
It is therefore sufficient to show that 
\begin{eqnarray}
\label{necessary to prove}
	f(
		\delta_{1}
	) \geq
	f(
	\delta_{2}
	)
	\Longleftrightarrow
	0 \leq \delta_{1} \leq \delta_{2} \leq \frac{1}{2}.
\end{eqnarray}
This equivalence will be shown in two steps:\\
First, we note that 
$
	0 \leq \delta_{1} \leq \delta_{2} \leq \frac{1}{2}
$
satisfy that $\forall\: 0 \leq \theta \leq \frac{1}{2}$:
\begin{eqnarray}
\label{f and delta equivalence 1}
	\delta_{1}\ln\left(
	 										\frac{\theta}{1-\theta}
							\right)
 \geq
	\delta_{2}\ln\left(
	 										\frac{\theta}{1-\theta}
							\right).
\end{eqnarray}
By adding 
$
	\ln\left(
				1-\theta
		\right)+
	{\xi E_{r}^{*}(\theta)}
$
to both sides of (\ref{f and delta equivalence 1}) we get:
\begin{eqnarray}
\label{f and delta equivalence 2}	
	\delta_{1}\ln\left(
	 										\frac{\theta}{1-\theta}
							\right)+
	\ln\left(
				1-\theta
		\right)+
	{\xi E_{r}^{*}(\theta)}
 \geq
	\delta_{2}\ln\left(
	 										\frac{\theta}{1-\theta}
							\right)+
	\ln\left(
				1-\theta
		\right)+
	{\xi E_{r}^{*}(\theta)}
\end{eqnarray}
or
\begin{eqnarray}
\label{f and delta equivalence 3}	
	\delta_{1}\ln{\theta}+
	\left(
			1-\delta_{1}
	\right)			
	\ln\left(
				1-\theta
		\right)+
	{\xi E_{r}^{*}(\theta)}		
 \geq
	\delta_{2}\ln{\theta}+
	\left(
			1-\delta_{2}
	\right)			
	\ln\left(
				1-\theta
		\right)+
	{\xi E_{r}^{*}(\theta)}.
\end{eqnarray}
This inequality is true for the values of $\theta$, which maximize the both sides of (\ref{f and delta equivalence 3}). i.e.:
\begin{eqnarray}
\label{f and delta equivalence 4}	
	\max_{0 \leq \theta \leq \frac{1}{2}}
	\left\{ 
		\delta_{1}\ln{\theta}+
		\left(
				1-\delta_{1}
		\right)			
		\ln\left(
					1-\theta
			\right)+
		{\xi E_{r}^{*}(\theta)}		
	\right\}		
 \geq
\nonumber\\
 	\max_{0 \leq \theta \leq \frac{1}{2}}
	\left\{ 
	\delta_{2}\ln{\theta}+
	\left(
			1-\delta_{2}
	\right)			
	\ln\left(
				1-\theta
		\right)+
	{\xi E_{r}^{*}(\theta)}		
	\right\} 	
\end{eqnarray}
or
\begin{eqnarray}
\label{f and delta equivalence 5}
	\max_{0 \leq \theta \leq \frac{1}{2}}
	f_\theta\left(
						\delta_{1}
					\right)
 \geq
 	\max_{0 \leq \theta \leq \frac{1}{2}}
	f_\theta\left(
						\delta_{2}
					\right).
\end{eqnarray}
In order to complete the proof, one must broaden the maximization ranges over $\theta$ in (\ref{f and delta equivalence 5}) into $0 \leq \theta \leq 1$. In order to justify that this broadening is possible, we present the following observation:\\
Each 
$
	0 \leq \delta \leq \frac{1}{2}
$
satisfy that
$
\forall \frac{1}{2} \leq \theta \leq 1
$:
\begin{eqnarray}
\label{maximization broaden - 0}
	\delta\ln\left(
	 										\frac{\theta}{1-\theta}
							\right)
 \leq
	\left(
		1-\delta
	\right)\ln\left(
	 										\frac{\theta}{1-\theta}
							\right).
\end{eqnarray}
By adding 
$
	\ln\left(
				1-\theta
		\right)+
	{\xi E_{r}^{*}(\theta)}
$
to both sides of (\ref{maximization broaden - 0}) we get:
\begin{eqnarray}
\label{maximization broaden - 1}
	\delta \ln\left(
	 										\frac{\theta}{1-\theta}
							\right)+
	\ln\left(
				1-\theta
		\right)+
	{\xi E_{r}^{*}(\theta)}
 \leq
	\left(
		1-\delta
	\right)\ln\left(
	 										\frac{\theta}{1-\theta}
							\right)+
	\ln\left(
				1-\theta
		\right)+
	{\xi E_{r}^{*}(\theta)}
\end{eqnarray}
or
\begin{eqnarray}
\label{maximization broaden - 2}
	\delta \ln{\theta}+
	\left(
			1-\delta
	\right)			
	\ln\left(
				1-\theta
		\right)+
	{\xi E_{r}^{*}(\theta)}		
 \leq
	\delta \ln\left(
				1-\theta
		\right)+
	\left(
		1-\delta
	\right)\ln{\theta}+
	{\xi E_{r}^{*}(\theta)}.
\end{eqnarray}
Using the fact that for the BSC model the ML error exponent, $E_{r}^{*}(\theta)$, is symmetric around $\theta=\frac{1}{2}$ (see (\ref{ML error_exponent - 2.5})), we can rewrite (\ref{maximization broaden - 2}) as:
\begin{eqnarray}
\label{maximization broaden - 3}
	\delta \ln{\theta}+
	\left(
			1-\delta
	\right)			
	\ln\left(
				1-\theta
		\right)+
	{\xi E_{r}^{*}(\theta)}		
 \leq
	\delta \ln\left(
				1-\theta
		\right)+
	\left(
		1-\delta
	\right)\ln{\theta}+
	{\xi E_{r}^{*}(1-\theta)}		
\end{eqnarray}
or
\begin{eqnarray}
\label{maximization broaden - 4}
	f_\theta\left(
						\delta(\bx,\by)
					\right)
 \leq
	f_{1-\theta}\left(
								\delta(\bx,\by)
							\right).
\end{eqnarray}
The meaning of (\ref{maximization broaden - 4}) is that when 
$
	0 \leq \delta \leq \frac{1}{2}
$, 
for each $\frac{1}{2} \leq \theta \leq 1$, 
$
f_\theta\left(
						\delta
					\right)
$
is always upper bounded by 
$
f_{1-\theta}\left(
								\delta
							\right) 
$
where $0 \leq 1-\theta \leq \frac{1}{2}$.
Thus, maximization of
$
f_\theta\left(
						\delta
					\right)
$
over $0 \leq \theta \leq 1$ is obviously accomplished by $\theta$ in $\left[0,\frac{1}{2}\right]$.\\
Therefore, (\ref{f and delta equivalence 5}) finally becomes:
\begin{eqnarray}
\label{f and delta equivalence updated 1}
	\max_{0 \leq \theta \leq 1}
	f_\theta\left(
						\delta_{1}
					\right)
 \geq
 	\max_{0 \leq \theta \leq 1}
	f_\theta\left(
						\delta_{2}
					\right)
\end{eqnarray}
thus,
\begin{eqnarray}
	0 \leq \delta_{1} \leq \delta_{2} \leq \frac{1}{2}
	& \Leftrightarrow &
	f\left(
			\delta_{1}
	\right)
 \geq
	f\left(
			\delta_{2}
	\right),
\end{eqnarray}
and the proof is complete.



%

\end{document}